\documentclass[journal,comsoc]{IEEEtran2}
\usepackage{todonotes}

\usepackage{cite}
\usepackage{amsmath,amssymb,amsfonts}
\usepackage{algorithmic}
\usepackage{graphicx}
\usepackage{textcomp}

\usepackage[caption=false]{subfig}
\usepackage[hypertexnames=false,linkcolor={1 0 0}]{hyperref} 
\usepackage[capitalise,nameinlink,noabbrev]{cleveref}
\hypersetup{
    colorlinks=true,
    linkcolor= black,
    citecolor=[rgb]{0.89, 0.0, 0.13},
    urlcolor=[rgb]{0.01, 0.28, 1.0}
}
\usepackage{booktabs}
\usepackage{float}
\usepackage{multirow}
\usepackage[ruled,vlined,linesnumbered,noresetcount]{algorithm2e}

\usepackage[normalem]{ulem}

\newcommand{\update}[1]{#1}
\newcommand{\remove}[1]{}

\SetKwInput{KwInput}{Input}                
\SetKwInput{KwOutput}{Output}              

\newcommand{\etal}{\textit{et~al.}}
\newcommand{\secondbest}[1]{{#1}$^{\diamond}$}
\newcommand{\linelabel}[1]{}
\begin{document}

\Crefname{figure}{Suppl. Figure}{Suppl. Figures}
\Crefname{table}{Suppl. Table}{Suppl. Tables}
\Crefname{section}{Suppl. Section}{Suppl. Sections}

\title{\textit{Semixup}: In- and Out-of-Manifold Regularization for Deep Semi-Supervised Knee Osteoarthritis Severity Grading from Plain Radiographs}
%
%
%

\author{Huy Hoang Nguyen,
        Simo Saarakkala,
        Matthew B.\ Blaschko,
        and Aleksei Tiulpin
\thanks{This work was supported in part by the strategic funding of the University of Oulu, in part by KAUTE  foundation, Finland and in part by Sigrid Juselius Foundation, Finland.}
\thanks{Huy Hoang Nguyen was with the Research Unit of Medical Imaging, Physics and Technology, University of Oulu, Finland. E-mail: huy.nguyen@oulu.fi.}
\thanks{Simo Saarakkala was with Research Unit of Medical Imaging, Physics and Technology, 
 University of Oulu, Finland and Department of Diagnostic Radiology, Oulu University Hospital, Finland. E-mail: simo.saarakkala@oulu.fi}
\thanks{Matthew B.\ Blaschko was with Center for Processing Speech \& Images, KU Leuven,  Belgium. E-mail: matthew.blaschko@esat.kuleuven.be.}
\thanks{Aleksei Tiulpin was with Research Unit of Medical Imaging, Physics and Technology, 
 University of Oulu, Finland,  Department of Diagnostic Radiology, Oulu University Hospital, Finland and Ailean Technologies Oy, Finland. E-mail: aleksei.tiulpin@oulu.fi.}
}

%
%

\markboth{IEEE Transactions on Medical Imaging, 2020}%
{Shell \MakeLowercase{\textit{et al.}}: Bare Demo of IEEEtran.cls for IEEE Communications Society Journals}

\maketitle

\begin{abstract}
\label{sc:abstract}
Knee osteoarthritis (OA) is one of the highest disability factors in the world\remove{ in humans}. This musculoskeletal disorder is assessed from clinical symptoms, and typically confirmed via radiographic assessment. This visual assessment done by a radiologist requires experience, and suffers from \linelabel{ln:abs_wording_1}\update{moderate to }high inter-observer variability. The recent \remove{development in the }literature has shown that deep learning \remove{(DL)} methods can reliably perform the OA severity assessment according to the gold standard Kellgren-Lawrence (KL) grading system. However, these methods require large amounts of labeled data, which are costly to obtain. In this study, we propose the \emph{Semixup} algorithm, a semi-supervised learning (SSL) approach to leverage unlabeled data. \emph{Semixup} relies on consistency regularization using in- and out-of-manifold samples, together with interpolated consistency. On an independent test set, our method significantly outperformed other state-of-the-art SSL methods in most cases\update{.}\remove{and even achieved a comparable performance to a well-tuned fully supervised learning (SL) model that required over 12 times more labeled data.} \linelabel{ln:data_efficient_edit} \update{Finally, when compared to a well-tuned fully supervised baseline that yielded a balanced accuracy (BA) of $70.9\pm0.8\%$ on the test set, \emph{Semixup} had comparable performance -- BA of $71\pm0.8\%$ ($p=0.368$) while requiring $6$ times less labeled data. These results show that our proposed SSL method allows building fully automatic OA severity assessment tools with datasets that are available outside research settings.}

\end{abstract}


%
\IEEEpeerreviewmaketitle

\section{Introduction}
\IEEEPARstart{O}{steoarthritis} (OA) is the most common joint disorder in the world causing enormous burdens at personal and societal levels~\cite{Hunter2019}. OA has an unknown etiology, and its indications at late stages are worn cartilage, bone deformity, and synovitis~\cite{dieppe2005pathogenesis, mobasheri2016update, mathiessen2017synovitis}. 

The most common joints affected by OA are knee and hip, and among these, the disease is more prevalent in knee~\cite{palazzo2016risk, ferket2017impact, cross2014global, SALIH2013482}. At the population level, such factors as sex, body-mass index (BMI), and age are known to be associated with OA~\cite{guccione1994effects, murtagh2004gender, glyn2015osteoarthritis}. As such, it was previously shown that people with BMI over $30$ have $7$-fold higher risk of knee OA than ones with BMI below 25~\cite{toivanen2009obesity}, and a half of elderly people over $65$ years of age have OA in at least one joint~\cite{miller2001modifiers}.

From an economic perspective, OA leads to a huge burden in terms of direct costs (e.g, hospitalization, diagnosis, and therapy), and indirect ones (e.g.\ losses of working days and productivity)~\cite{leardini2004direct}. For example, in the United States, OA costs hundreds of billion dollars annually, and is in the top-5 of annual Europe healthcare expenditure~\cite{mobasheri2016update, cross2014global}.

Currently, knee OA diagnosis starts with a clinical examination, and then, a radiographic confirmation takes place when necessary~\cite{palazzo2016risk, HUNTER20191745}. However, such\linelabel{ln:intro_wording_1}\remove{guideline} \update{practice} enables knee OA diagnosis only at a late stage when the cartilage is already worn, and the bone deformity is present, which leads to severe pain, and even physical disability~\cite{dieppe2005pathogenesis, mobasheri2016update}. Ultimately, the only remaining option for a patient in that scenario is total knee replacement (TKR) surgery.

The literature shows a large and rapidly growing number of TKR surgeries worldwide~\cite{CARR20121331, PRICE20181672, SALIH2013482}. As such, the annual rate of TKR surgeries in the United States has doubled since $2000$ for adults of $45$-$64$ years old~\cite{baker2013influence, ferket2017impact}. Therefore, there is a need for prevention of global disability.

\begin{figure*}[htbp]
    \centering
        \subfloat[KL $0$ \label{fig:kl0}]{\includegraphics[scale=0.45]{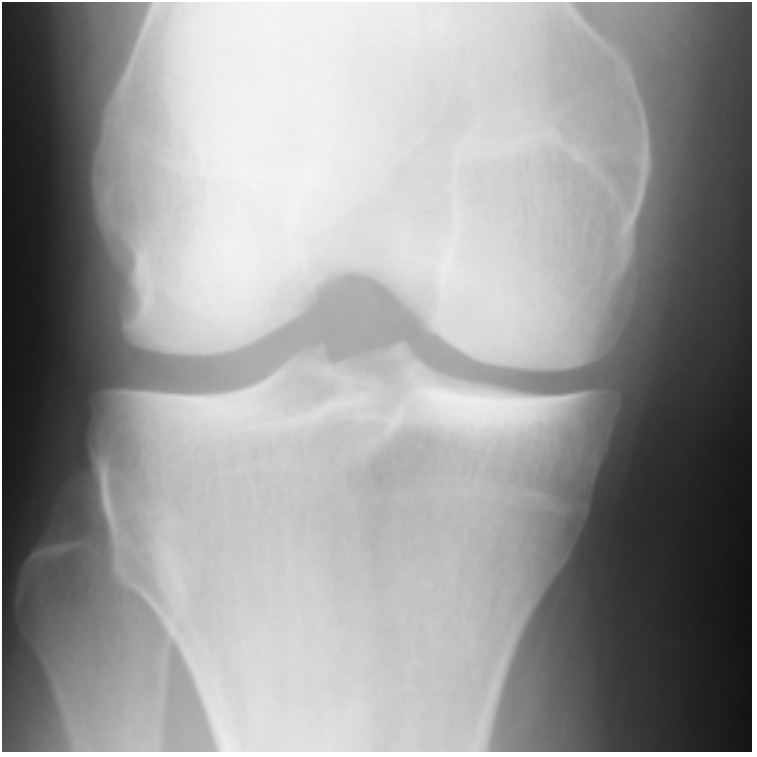}}\hfill%
        \subfloat[KL $1$ \label{fig:kl1}]{\includegraphics[scale=0.45]{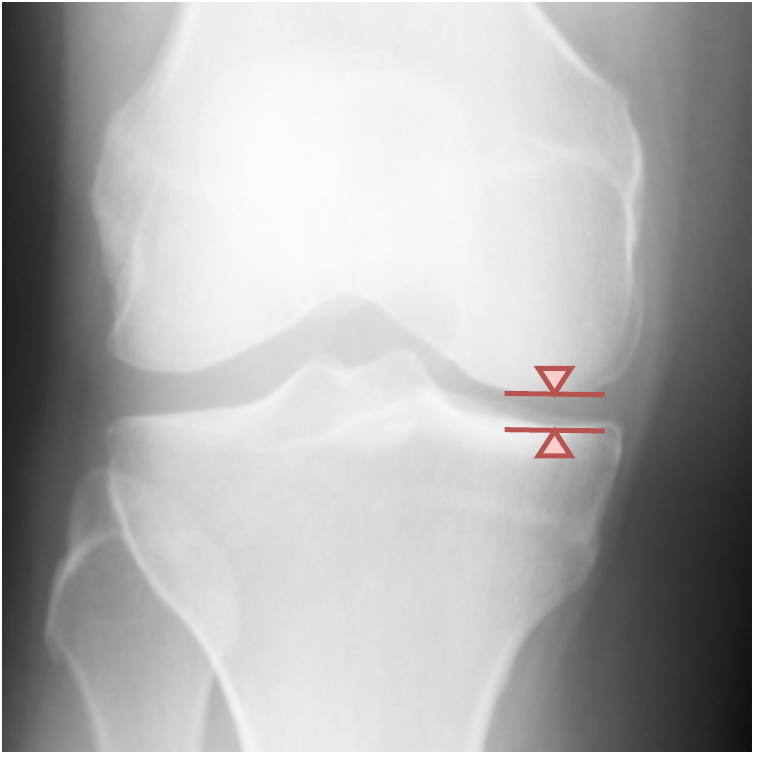}}\hfill%
        \subfloat[KL $2$ \label{fig:kl2}]{\includegraphics[scale=0.45]{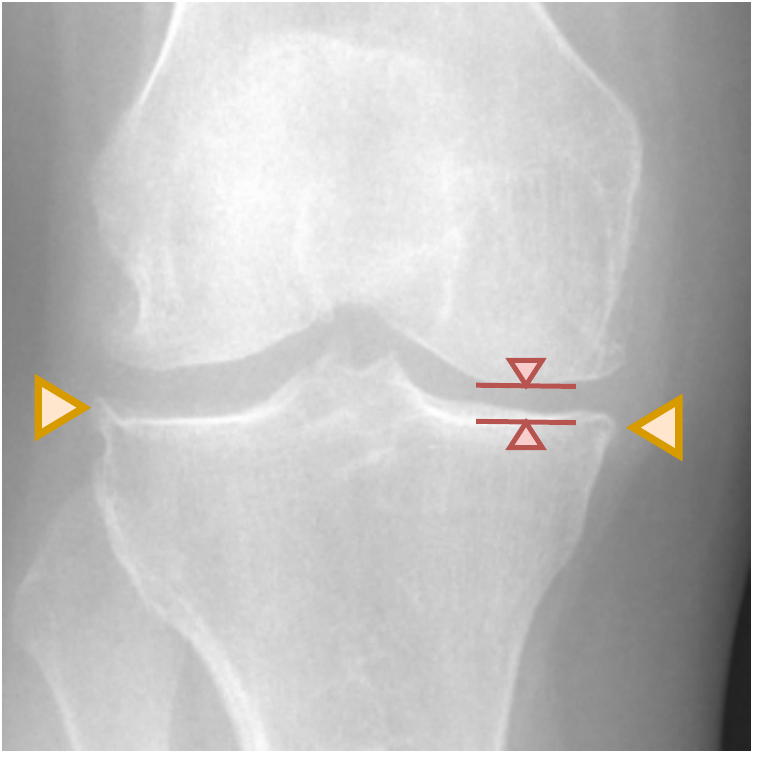}}\hfill%
        \subfloat[KL $3$ \label{fig:kl3}]{\includegraphics[scale=0.45]{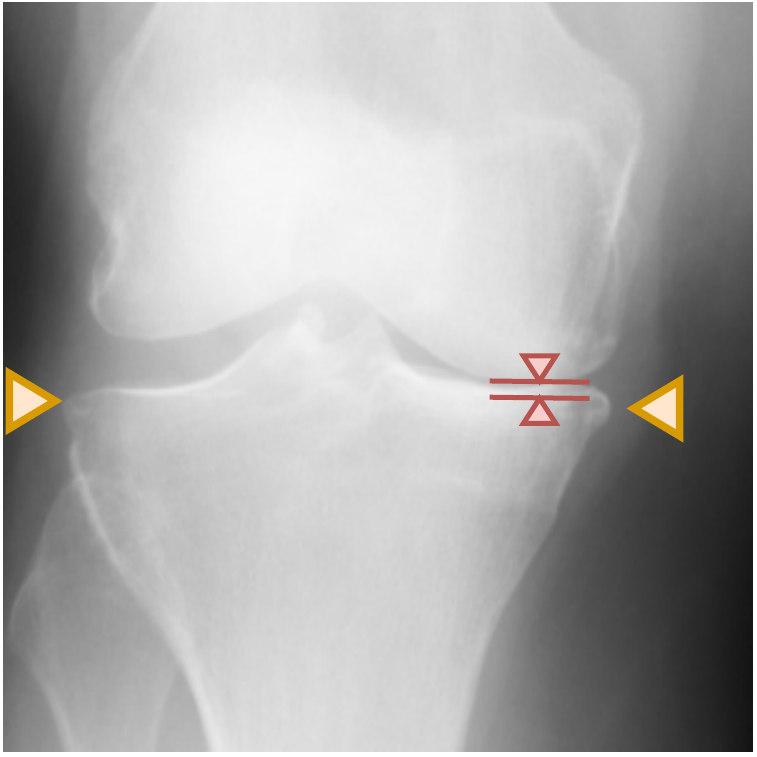}}\hfill%
        \subfloat[KL $4$ \label{fig:kl4}]{\includegraphics[scale=0.45]{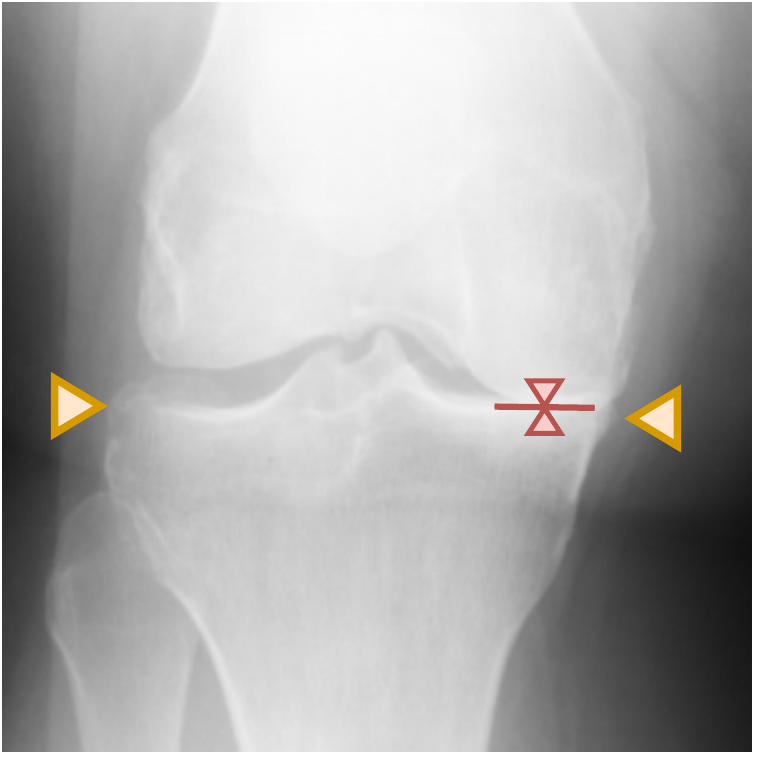}}\hfill%
    \caption{\small Samples of knee radiographs. Joint space narrowing and osteophyte features are indicated by \remove{blue squares and red circles}\update{red and yellow marks} respectively. (a) KL 0: A healthy knee without OA, (b) KL 1 (Doubtful OA): Potential joint space narrowing, (c) KL 2 (Mild OA): Clear evidences of osteophytes, as well as slight reduction of joint space, (d) KL 3 (Moderate OA): Osteophytes grow and joint space narrowing progresses badly, and (e) KL 4 (Severe OA): Besides osteophytes, joint space is reduced so severely that the tibia and the femur are connected.}
    \label{fig:samples_by_kl}
\end{figure*}

Imaging, in contrast to clinical examination, may enable the detection of early knee OA signs at the stages when behavioral interventions (e.g. exercises and weight loss programs) could slow down the disease progression~\cite{baker2000exercise}. Radiographic assessment is the foremost imaging tool for detecting knee OA in primary care, and Kellgren-Lawrence (KL) is one of the most common clinical scales for the assessment of OA severity from plain radiographs (\cref{fig:samples_by_kl}). However, visual diagnosis done by a radiologist suffers from low inter-rater agreement~\cite{reichenbach2008prevalence, tiulpin2018automatic}, thereby introducing large inconsistencies into decision-making. One possible solution to make OA diagnosis more systematic and allow for the detection of knee OA at early stages is to leverage computer-aided methods for image analysis~\cite{buckland1986quantitative}. Deep learning (DL) has become a state-of-the-art approach in this realm, and recent 
studies~\cite{antony2017automatic, antony2018automatic, tiulpin2018automatic} have demonstrated that DL-based methods allow \remove{for }fully-automatic KL grading. Furthermore, these studies showed a high level of agreement between the predictions made by DL-based models and the annotations produced by a consensus of radiologists. 

Despite good and promising results, all the previously published DL methods in OA domain were based on Supervised-Learning (SL) and required large amounts of labeled data, which are not currently widely available. 
In practical applications, such datasets as the Osteoarthritis Initiative (OAI, \url{https://nda.nih.gov/oai/}) and the Multicenter Osteoarthritis Study (MOST, \url{http://most.ucsf.edu/}) are expensive to obtain due to high costs of data \linelabel{ln:intro_wording_2}\remove{collection}\update{acquisition} and annotation. As such, the latter needs multiple skilled experts (e.g.\ radiologists or orthopedists) making the process even more costly. Whereas labeled data are difficult to acquire, unlabeled data are available in large amounts, and can be collected from hospital imaging archives at low cost. 

In the natural image recognition domain, it has been shown that leveraging small amounts of labeled\remove{,} and large amounts of unlabeled data in a Semi-Supervised Learning (SSL) setting could potentially resolve the need for large labeled datasets. Recent studies~\cite{laine2016temporal, verma2019interpolation, berthelot2019mixmatch} have developed SSL methods to utilize unlabeled data during the training processes, and have achieved competitive performances in image classification benchmarks using only a small fraction of the labeled data used in fully supervised settings. 

Many SSL-based applications in the medical domain have previously been developed for automatic disease diagnosis. However, most of those are related to medical image segmentation~\cite{su2016interactive, wang20144dactivecut, iglesias2010agreement, kumar2019hyperspectral, li2018semi}, and use generative adversarial networks (GANs)~\cite{goodfellow2014gan} as a core method. To the best of our knowledge, there have been no SSL-based methods developed in the knee OA realm.

In this study, we, for the first time in the OA field, propose to leverage SSL for automatic assessment of knee OA severity from plain radiographs. Inspired by previous research in SSL~\cite{laine2016temporal, verma2019interpolation, belkin2006manifold}, and the recently developed technique \textit{mixup}~\cite{zhang2017mixup}, we propose a novel SSL method -- \emph{Semixup}, providing its systematic empirical comparison with the state-of-the-art approaches. Specifically, our contributions are the following:
\begin{enumerate}
    \item We enhance the state-of-the-art supervised baseline~\cite{tiulpin2018automatic}, and propose a novel \textit{Separable Adaptive Max-pooling} (SAM), as a drop-in replacement for the Global Average Pooling (GAP). This allows us to significantly improve over previously reported supervised results. 
    \item We introduce a novel semi-supervised DL-based method called \emph{Semixup} for automatic KL grading of knee OA from plain radiographs. Our method yields competitive results to a well-tuned SL model trained on over $6$ times more labeled data.
    \item We systematically compare our method against several state-of-the-art SSL methods, and experimentally show that \textit{Semixup} outperforms them in nearly all data regimes.
    \item We follow the guidelines from~\cite{oliver2018realistic} to conduct a realistic evaluation of our SSL approach, and provide insights into the scalability of our method with respect to the number of labeled examples, together with a tractable amount of unlabeled samples.
\end{enumerate}

\begin{figure*}[!ht]
    \centering
    \IfFileExists{figs/semixup/semixup_workflow-crop.pdf}{}{\immediate\write18{pdfcrop figs/semixup/semixup_workflow.pdf}}
    \IfFileExists{figs/semixup-illustration-crop.pdf}{}{\immediate\write18{pdfcrop figs/semixup-illustration.pdf}}
    \subfloat[Workflow of unsupervised branch.  \label{fig:workflow}]{\includegraphics[scale=0.65]{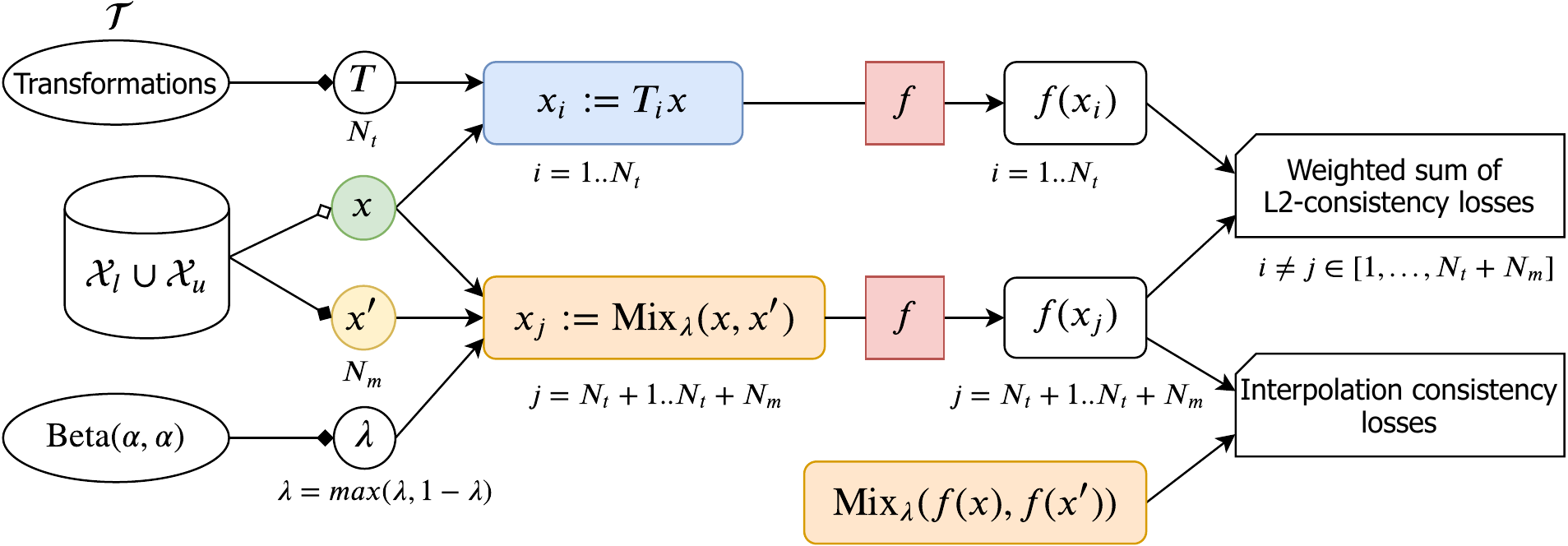}}\hfill%
    \subfloat[Illustration \label{fig:illustration}]{\includegraphics[scale=0.65]{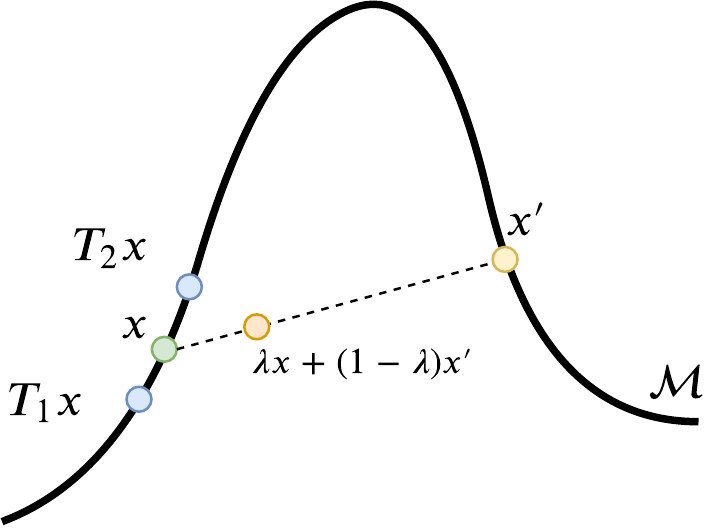}}\hfill%
    \caption{\small The main idea of \textit{Semixup}: (\protect\subref*{fig:workflow}) The workflow of its in- and out-of-manifold consistency regularization branch, and (\protect\subref*{fig:illustration}) its conceptual illustration. For each image $x \in L \cup U$, we sample $N_t$ transformations, $N_m$ images $x'$, and \textit{mixup} coefficients $\lambda$ (s.t. $\lambda > 0.5$) to transform and blend with $x$. Then, we enforce consistency of predictions for every pair $x_i, x_j$ under L2 norm. In addition, we use the interpolation consistency regularizer. Links with $\blacksquare$ denote sampling with replacement, ones with $\square$ indicate sampling without replacement.}
    \label{fig:mainflow}
\end{figure*}
\section{Related Work}

\subsection{Deep Semi-Supervised Learning}
There exists a wide variety of DL-based SSL methods in the literature; however, we discuss here only the ones that are close to our method and yield state-of-the-art results on generic image recognition datasets. Such approaches use two main ideas: \textit{consistency regularization} and \textit{pseudo-labeling}. 

Consistency regularization is based on the assumption that the predictions of the model $P(y|\mathbf{x})$ and $P(y|T\mathbf{x})$, where $\mathbf{x}$ is a data point and $T$ -- class-preserving stochastic data augmentation -- should not differ. The methods using the technique applied to unlabeled data include the $\Pi$-model~\cite{laine2016temporal}, and Mean Teacher (MT)~\cite{tarvainen2017mean}.

Label guessing (or pseudo-labeling)~\cite{Zhu02learningfrom} was proposed in a deep learning setting in~\cite{lee2013pseudo}, and uses predicted labels for unlabeled samples with high confidences to update the gradients of the neural network. This technique can also be viewed as entropy regularization which favors low-density separation between classes~\cite{chapelle2005semi}. 

The aforementioned SSL techniques have been explored separately; however, Berthelot~\etal\ has recently introduced the MixMatch method~\cite{berthelot2019mixmatch}, the idea of which was to combine label guessing and entropy regularization into a holistic framework. The authors of MixMatch made an empirical observation that applying \textit{mixup}~\cite{zhang2017mixup}, which performed convex combinations of 2 arbitrary input data points $x$ and $x'$, and their labels $y$ and $y'$ (i.e.\ $\lambda x + (1-\lambda) x'$ and $\lambda y + (1-\lambda) y'$, where $\lambda \sim \mathrm{Beta}(\alpha, \alpha)$, for $\alpha \in (0, \infty)$), to labeled data and unlabeled data with guessed labels helps to improve the performance. In our method, we also use \textit{mixup} for both labeled and unlabeled data; however, we avoid using label guessing due to its potential to propagate label errors that can be common in medical domains. 

We finally discuss here two main limitations of all the mentioned recent studies on SSL. The first common limitation of those methods is that their evaluations were done on major image recognition benchmarks, namely CIFAR-10 and/or ImageNet without considering real-world problems, such as the ones related to medical image recognition.

\linelabel{ln:intro_limitation}\update{The second limitation of the aforementioned studies is that the  methods proposed in them were evaluated only with respect to the change of the amount of labeled data. However, none of those studies explored the amount of unlabeled data needed to improve the performance. In our work, we conduct such evaluation.}


\subsection{Deep Learning for Knee Osteoarthritis Diagnosis}
Fully supervised DL-based methods that use Convolutional Neural Networks (CNN) have recently been used to assess knee OA severity. In particular, in their pioneering work, Antony~\etal~\cite{antony2016quantifying} applied transfer learning and showed significant improvements compared to classifiers trained using hand-crafted features. In the follow-up work~\cite{antony2017automatic}, Antony~\etal\ proposed a CNN architecture trained from scratch to classify knee images according to the KL scale. Their new approach outperformed their previous transfer learning results. 

The main common limitation of the aforementioned studies was that neither of them utilized an independent test set. In contrast, Tiulpin~\etal~\cite{tiulpin2018automatic} addressed this limitation and also proposed a novel CNN architecture that outperformed the previous methods~\cite{antony2016quantifying, antony2017automatic}. Interestingly, that model performed on-par with transfer learning baseline while having significantly less trainable parameters thanks to the inductive bias from using the relative symmetry of visual features in knee images. 

\linelabel{ln:rw_wording_1}\remove{The}\update{More} recent studies by Chen~\etal~\cite{chen2019fully}, Norman ~\etal~\cite{norman2019applying}, and G{\'o}rriz~\etal~\cite{gorriz2019assessing} did not use any independent test set either. The only latest study where an independent test set was used for assessment of the results was by Tiulpin~\etal~\cite{tiulpin2019automatic}. In that study, the authors obtained a KL classification model as a bi-product of their main method; however, they obtained results that are similar to their previous study~\cite{tiulpin2018automatic}. 

Despite having an independent test set and state-of-the-art results, Tiulpin~\etal~\cite{tiulpin2018automatic, tiulpin2019automatic} used large amounts of labeled data for training. We emphasize here that none of the existing studies in which DL was applied for knee OA diagnosis from radiographs addressed the question of the quality of automatic KL grading as a function of the dataset size. Our work answers this question via a thorough experimental evaluation of both SL and SSL methods.

\section{Method}
\subsection{Overview}
The method proposed in this paper consists of two parts: 1) a novel extension of a previously developed Siamese network developed by Tiulpin~\etal~\cite{tiulpin2018automatic} and 2) a novel deep SSL technique. The utilized Siamese model uses shared branches of the CNN which focus their attention on the medial and the lateral sides of the analyzed knee (see \cref{fig:architecture}). Here, we consider pairs of lateral and medial image patches as single data points $x\in \mathbb{R}^{2\times H \times H}$, where $H$ is the size of the image patch. KL grades are the outputs of our model: $y\in \left\{0,1,2,3,4\right\}$. $f_\theta$ denotes a Siamese neural network with parameters $\mathbf{\theta}$. In our setting, $p(y|\mathbf{x})=f_{\mathbf{\theta}}(\mathbf{x})$.

Our SSL method aims to perform penalization of local sharpness of the surface loss along the data manifold $\mathcal{M}$ and also within its surroundings. The former is achieved via minimization of $\mathbb{E}_x{\|J_\mathcal{M}\|^2_F}$, where $J_\mathcal{M}$ denotes the Jacobian along the data manifold $\mathcal{M}$ and $\| \cdot \|_F$ is the Frobenius norm, as well as $\mathbb{E}_x{\|J_\theta\|^2_F}$, where $J_\theta$ denotes the Jacobian in the parameter space, using consistency regularization~\cite{laine2016temporal,athiwaratkun2018there}. To generate out-of-manifold samples, we use \textit{mixup}~\cite{zhang2017mixup}. Here, we first enforce linear behavior of the model along the \textit{mixup} rays via interpolation consistency training (ICT)~\cite{verma2019interpolation}, and then apply the aforementioned consistency regularization to enforce consistent behavior of the model along the \textit{mixup} rays which are in the close surroundings of $\mathcal{M}$. As our idea for SSL centers around \textit{mixup}, we name our method \textit{Semixup}. 
We graphically illustrate the process of sample generation for consistency regularization in \cref{fig:mainflow}.


\subsection{Network Architecture} \label{sc:architecture}
\begin{figure}[ht!]
    \centering
    \IfFileExists{figs/semixup/architecture-crop.pdf}{}{\immediate\write18{pdfcrop figs/semixup/architecture.pdf}}
        \includegraphics[scale=0.36]{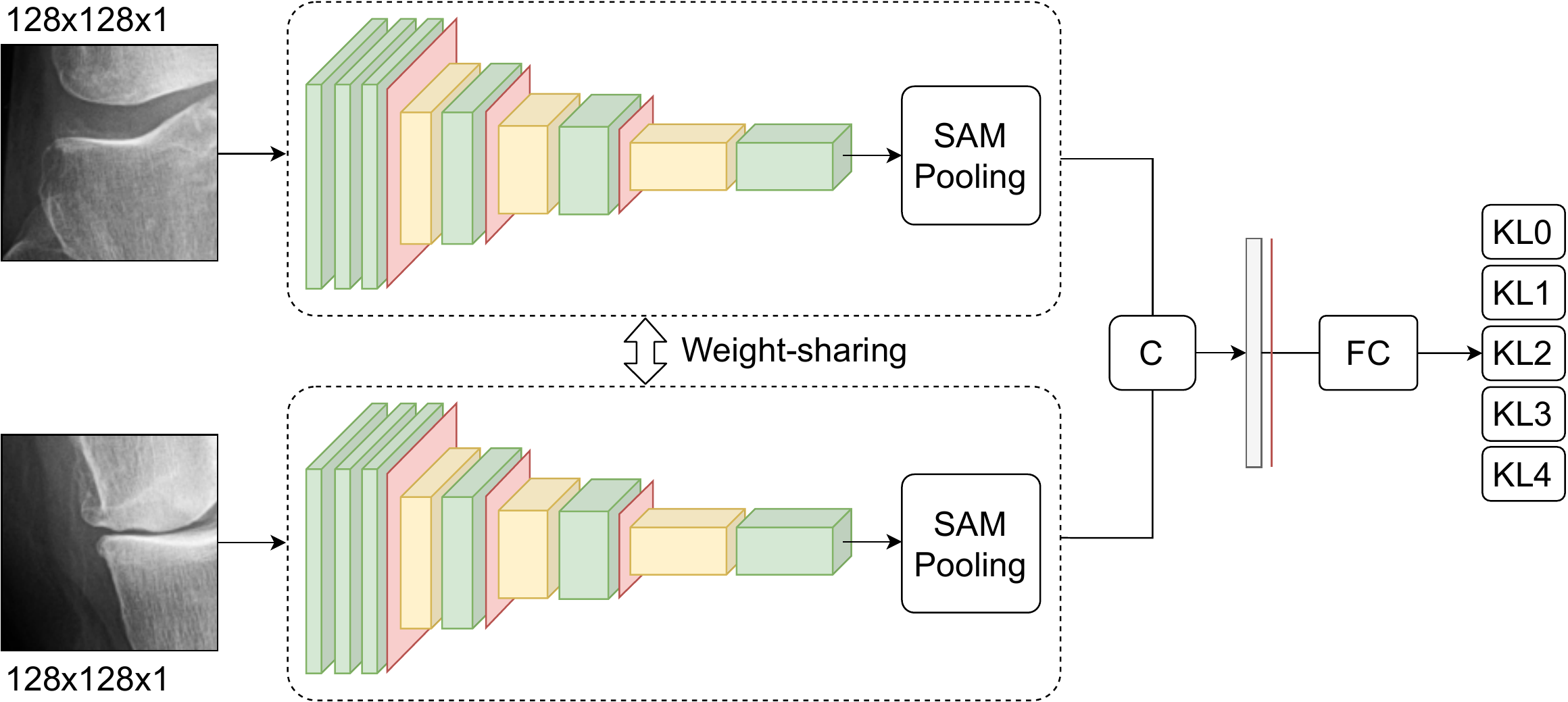}
    \caption{\small The Siamese architecture of our model. Green denotes the blocks that use $3\times3$ convolutions with the stride of 1, and yellow denotes $3\times3$ convolutions with the stride of 2. Red indicates dropout layers. After pooling the features by Separable Adaptive Max-pooling (SAM) and concatenating (C) them into a single vector, they are passed into a fully-connected (FC) layer that predicts KL grades.}
    \label{fig:architecture}
\end{figure}

We follow the design of the previously developed Siamese model by Tiulpin~\etal~\cite{tiulpin2018automatic}, and propose several modifications essential to obtain better KL grading performance in both SL and SSL settings. The schematic illustration of our proposed network is presented in \cref{fig:architecture}. 

The basic building block of our model consists of a $3\times 3$ convolution with a zero padding $P$ and a stride $S$, an instance normalization (IN), and a leaky rectified linear unit (LeakyReLU) activation with a slope of $0.2$. 

To enrich the representation power of our model, we first start with three consecutive $3\times3, S=1, P=1$, and one $3\times3, S=2, P=1$ convolutional blocks (green blocks in \cref{fig:architecture}). Subsequently, we alternate $3\times3, S=2, P=1$ (yellow blocks in \cref{fig:architecture}), and $3\times3, S=1, P=1$ blocks until a feature map of size $\frac{1}{8}H\times \frac{1}{8}H$ is obtained.

We use stridden convolutions to perform the downsampling, whereas the original model from~\cite{tiulpin2018automatic} used max-pooling layers. 
The main motivation for our method to use stridden convolutions is that the translation invariance achieved by the use of max-pooling could potentially harm the results by removing the dependencies between the OA-related fine-grained features at higher layers of the network~\cite{springenberg2014striving}. 

Similar issues could also arise in the bottleneck of the network, where Tiulpin~\etal~\cite{tiulpin2018automatic} used Global Average Pooling (GAP).
We argue that averaging such large feature maps is not the most optimal pooling strategy, and we tackle this problem via our novel SAM pooling (see \cref{sc:sam}).

All the blocks described above share the weights among the branches of our Siamese CNN, where each branch of the model processes an individual side of the knee image pair (lateral or medial). \linelabel{ln:sam_output}\update{Similar as GAP, SAM layer in each branch produces}\remove{Each branch of this model generates} $1\times1$ features, which are concatenated, passed through a dropout, and subsequently fed into a fully-connected layer the OA severity stage. A detailed description of our architecture is provided in \Cref{tbl:detailed_arch}.

\subsection{Separable Adaptive Max-pooling} \label{sc:sam}
As mentioned previously, we propose a replacement for GAP to deal with the potential information loss in the bottleneck of the model. \linelabel{ln:sam_clarification} The proposed SAM scheme is based on the idea of firstly applying pooling along one direction of the feature map (horizontal or vertical). Secondly, we use a $1\times 1$ convolutional block (with IN and LeakyReLU) to remove unnecessary correlations between the pooled features. Finally, we apply the pooling in the \linelabel{ln:sam_edit}\remove{other} direction \update{orthogonal to the one} \remove{than was done} in the first phase. 

\begin{figure}[!t]
    \centering
    \IfFileExists{figs/pooling/sam-h-crop.pdf}{}{\immediate\write18{pdfcrop figs/pooling/sam-h.pdf}}
    \IfFileExists{figs/pooling/sam-v-crop.pdf}{}{\immediate\write18{pdfcrop figs/pooling/sam-v.pdf}}
    
        \hspace*{\fill}%
        \subfloat[SAM-VH \label{fig_sam_vh}]{\includegraphics[scale=0.4]{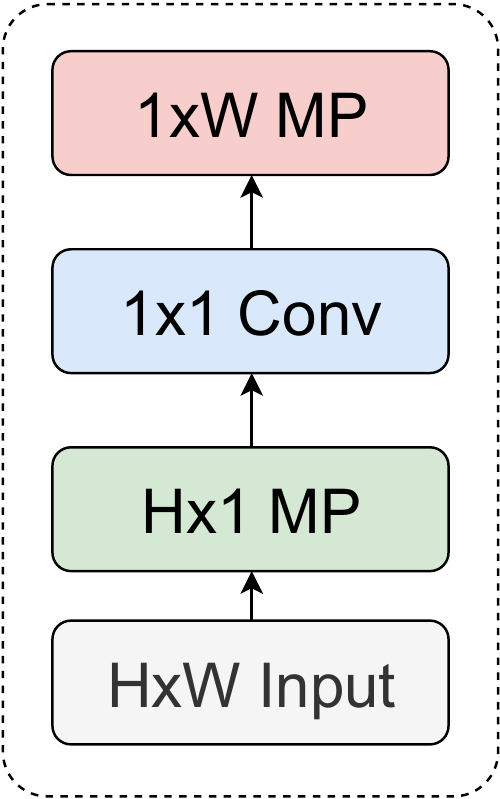}
        }\hfill%
        \subfloat[SAM-HV 
        \label{fig_sam_hv}]{\includegraphics[scale=0.4]{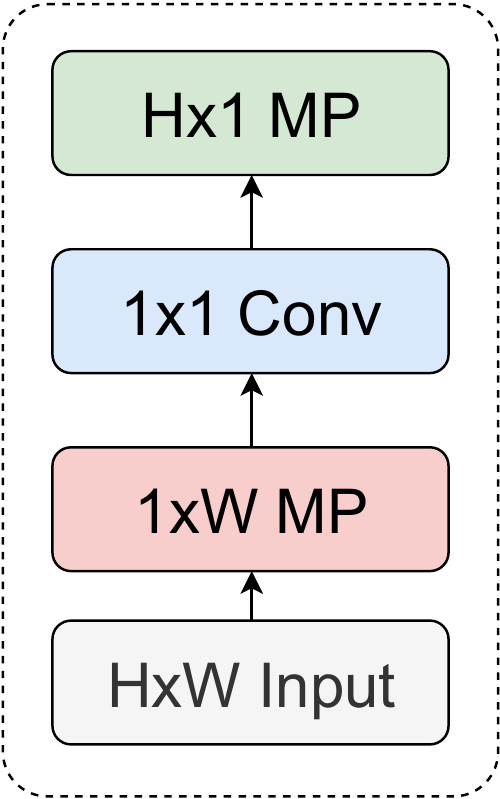}
        }\hspace*{\fill}%
    \caption{\small \emph{Separable Adaptive Max-pooling} configurations. The core idea of this approach is to inject a non-linearity between the pooling steps -- vertical-horizontal and horizontal-vertical as displayed in subplots (\protect\subref*{fig_sam_vh}) and (\protect\subref*{fig_sam_hv}), respectively. Here, H and W indicate the height and the width of the input, respectively.
    }
    \label{fig:sam_all}
\end{figure}

As the initial pooling of the features can be done in either horizontal or vertical directions, we suggest two configurations of SAM presented in \cref{fig:sam_all}, namely:

\begin{enumerate}
    \item \linelabel{ln:sam}Max-pooling $\frac{1}{8}H\times 1$ with stride $\frac{1}{8}H\times 1$, $1\times 1$ convolution, IN, LeakyReLU with the slope of $0.2$, and $1\times \frac{1}{8}H$ max-pooling with the stride of $1\times 1$ (SAM-VH).
    \item Similar to the above, but the first and the last max-poolings are swapped (SAM-HV).
\end{enumerate}

\subsection{Semi-Supervised Learning}
\subsubsection{Problem setting}
Let $\mathcal{X}_l$ and $\mathcal{X}_u$ be labeled and unlabeled image sets, respectively. Let $\mathcal{Y}$ denote the labels for $\mathcal{X}_l$. In our setting, we optimize the following objective:

\begin{equation}\label{eqn:loss}
    \underset{\theta}{\mathrm{min}}\ \mathcal{L}_{l}(\theta; \mathcal{X}_l, \mathcal{Y}) + \mathcal{L}_u(\theta;\mathbf{w}, \mathcal{X}_u, \mathcal{X}_l),
\end{equation}
where $\mathcal{L}_{l}$ is a cross-entropy loss with \textit{mixup}, $\mathcal{L}_u$ is a combination of losses without the involvement of labels, and $\mathbf{w}$ are hyperparameters responsible for weighing unsupervised losses. Here, $\mathcal{L}_u$ acts as a regularizer which leverages data without labels to enhance the robustness of the model $f_{\theta}$ via auxiliary tasks.

\subsubsection{Supervised Loss}
\textit{mixup} proposed in~\cite{zhang2017mixup} is a simple and effective technique to improve 
generalization. It can be viewed as data augmentation, and in a nutshell, it linearly mixes two samples $x_i$, $x_j$  with a blending coefficient $\lambda \sim \textrm{Beta}(\alpha,\alpha)$, for $\alpha \in (0, \infty)$: 
\begin{align} \label{eqn:mixup}
x_{mix} = \textrm{Mix}_\lambda(x_i, x_j) = \lambda x_i + (1-\lambda) x_j .
\end{align}

Having the mixed sample $x_{mix}$, the following loss is optimized:
\begin{multline}
    \mathcal{L}_l(\theta; x_{mix}, y_i, y_j) = \lambda \mathcal{L}_{ce}\left(x_{mix}, y_i\right) +  \\
    + (1 - \lambda) \mathcal{L}_{ce}(x_{mix}, y_j),
    \label{eqn:soft_ce}
\end{multline}
where $ \mathcal{L}_{ce}$ is a multi-class cross-entropy loss. Here and further, we call the loss in \eqref{eqn:soft_ce} "soft" cross-entropy loss.

After the introduction of~\textit{mixup}, it was shown that this technique performs out-of-manifold regularization~\cite{guo2019mixup}, and in our work, we exploit this property. Specifically, we view \textit{mixup} as a data augmentation technique that generates out-of-manifold samples $x_{mix}$ that belong to the ray between any two points $x_i$ and $ x_j$ (\cref{eqn:mixup}, and \cref{fig:illustration}).

\subsubsection{Consistency Regularization: Penalizing Local Loss Sharpness} \label{sc:consistency_regularization}
The consistency regularization technique used in~\cite{laine2016temporal, tarvainen2017mean} aims to minimize the following objective w.r.t $\theta$:
\begin{equation}\label{eqn:consistency}
    \mathbb{E}_{x \sim p(x)}\mathbb{E}_{T, T' \sim p(\tau)} \left \| f_{\theta}(Tx) - f_{\theta}(T'x) \right \|_{2}^{2},
\end{equation}
where $p(x)$ is the distribution of training data (labeled and unlabeled, and $p(\tau)$ is the distribution of stochastic transformations (e.g., data augmentations). 

According to Athiwaratkun~\etal~\cite{athiwaratkun2018there}, regularizing consistency for a model using dropout in convolutional layers implies minimization of two terms: $\mathbb{E}_x{\|J_\mathcal{M}\|^2_F}$ and $\mathbb{E}_x{\|J_\theta\|^2_F}$. 
That idea results in an interesting connection between the consistency-based method and the classic graph-based approaches that use Laplacian regularization for SSL~\cite{chapelle2009semi}. 

The explanations provided by Athiwaratkun~\etal~\cite{athiwaratkun2018there} demonstrate that minimization of $\mathbb{E}_x{\|J_\theta\|^2_F}$ leads to a broader optimum that is presumably helpful for good model generalization. Therefore, the minimization of the regularization term from \eqref{eqn:consistency} can lead to better performance which has been supported by experimental evidence in~\cite{laine2016temporal, tarvainen2017mean, athiwaratkun2018there}.

\subsubsection{Interpolation Consistency}
The \textit{mixup} operator was introduced as an efficient data augmentation for supervised learning regularizing against out-of-manifold samples that lie close to $\mathcal{M}$. Recently, Verma~\etal~\cite{verma2019interpolation} utilized it to enforce linear behavior of the model along the \textit{mixup} rays in the ICT method:
\begin{equation}\label{eqn:ict}
    \mathbb{E}_{\lambda \sim \textrm{Beta}(\alpha, \alpha)}\mathbb{E}_{x_i, x_j \sim p(x)} \left \| \textrm{Mix}_\lambda(f_{\theta}(x_{i}), f_{\theta}(x_{j})) - f_{\theta}(x_{mix}) \right \|_{2}^{2}.
\end{equation}
While Verma~\etal~\cite{verma2019interpolation} did not consider \textit{mixup} as an out-of-manifold regularizer, we consider this to be an important observation~\cite{guo2019mixup}. 
 
\subsection{Semixup} \label{sc:semixup}

\subsubsection{Motivation}
The consistency regularization and ICT aim to maximize the consistency of label assignment within and out of $\mathcal{M}$. However, we also note that while linear behavior of the model is achieved in ICT, it does not aim to minimize the inconsistency of label assignment for e.g. $Tx$ and $x_{mix}$ which can be viewed as in- and out-of-manifold augmented versions of a data point $x$. In this work, we address this limitation and strengthen the ability of making consistent predictions for out-of-manifold samples that are close to the data manifold $\mathcal{M}$. In summary, besides applying loss \eqref{eqn:soft_ce} over labeled data, \textit{Semixup} optimizes a linear combination of the objectives shown in \eqref{eqn:consistency}, and~\eqref{eqn:ict}, as well as the out-of-manifold term described in the following section over both labeled and unlabeled samples. Supplementary Algorithm \ref{alg:semixup} shows a concrete implementation of our method.

\subsubsection{Out-of-Manifold Consistency Regularization}
We use the aforementioned regularizers from \eqref{eqn:consistency} and \eqref{eqn:ict}. Subsequently, given the motivation above, we minimize the following additional objective: 
\begin{equation}\label{eqn:semixup_term}
    \mathbb{E}_{\lambda \sim \textrm{Beta}(\alpha, \alpha)}\mathbb{E}_{x ,x' \sim p(x)}\mathbb{E}_{T \sim p(\tau)}\left \|f_{\theta}(\textrm{Mix}_{\lambda}(x, x'))-f_{\theta}(Tx)\right \|_{2}^{2}.
\end{equation}

The presented objective aims to maximize the consistent label assignment for perturbed data items $Tx\in \mathcal{M}$, and also the ones being out-of-manifold but are close to it. To generate the latter, when sampling $\lambda$ in \textit{mixup}, we enforce $\lambda=\max(\lambda, 1-\lambda)$.

\subsubsection{Low-variance Sampling}
Our method 
uses both labeled and unlabeled data in the unsupervised regularization term of the loss \eqref{eqn:loss}. It is motivated by the fact that unlabeled data in medical imaging can come from different device vendors rather than labeled data, thereby the empirical distributions of labeled and unlabeled data might be misaligned. 

We note that the objective in \eqref{eqn:semixup_term} uses stochastic augmentations $T\sim p(\tau)$. When using this loss in a combination with consistency regularization and ICT, it can be seen that there exist predictions for two versions of an image $x$ as in \eqref{eqn:consistency}. Therefore, when optimizing with stochastic gradient descent, we are able to obtain a lower variance stochastic estimate of the objective~\eqref{eqn:semixup_term} at low marginal computational cost (see lines \ref{algo_line:semixup_1}, \ref{algo_line:semixup_2} of Supplementary Algorithm \ref{alg:semixup}).



\section{Experiments}
\subsection{Datasets}
\label{sc:datasets}
We used knee radiographs from two large public cohorts: The OAI and the MOST. The OAI dataset was collected from $4,796$ participants whose ages were from $45$ to $79$ years old. The cohort included a baseline, and follow-up visits after $12$, $18$, $24$, $30$, $36$, $48$, $60$, $72$, $84$, $96$, $108$, $120$, and $132$ months. Radiographic imaging for bilateral fixed flexion X-ray images took place at most, but not all follow-ups. We used \linelabel{ln:oai_data_filter} \update{the data from all the knees without implants that had KL grades available in the metadata} except for those imaged during $18$, $30$, \update{$48$}, $60$, $84$, $108$, \update{$120$, and $132$}-month follow-ups, where \update{knee radiograph} imaging was performed only for small sub-cohorts. \linelabel{ln:data_exclusion_footnote}\update{We graphically depict the OAI data selection in \Cref{fig:oai_splitting_workflow}}. The MOST dataset had $3,021$ participants examined at its baseline, and follow-up visits after $15$, $30$, $60$, $72$, and $84$ months. Except for the $72$-month follow-up, each examination included radiographic imaging.


Radiographs in the OAI dataset were posterior-anterior (PA) bilateral images acquired with the protocol that called for a beam angle of $10$ degrees. We utilized OAI data in training and model selection phases. 

The MOST dataset included PA, and lateral images. The PA images were acquired with a beam angle of $5$, $10$, or $15$ degrees. In this study, we used the MOST dataset as an independent test set. To ensure the reliability of the labels in MOST, we only \linelabel{ln:exp_typo_1}\remove{keep}\update{kept} $10$ degree bilateral PA images that were acquired at baseline, had KL and Osteoarthritis Research Society International (OARSI) grades available, and were without implants. Eventually, our training and test data comprised $39,902$ and $3,445$ knee images from the OAI, and the MOST datasets, respectively. The detailed data distribution by KL grade is presented in \cref{tbl:data_desc}.

\begin{table}[ht!]
\centering
\caption{\small Description of the OAI and the MOST datasets.}
\label{tbl:data_desc}
\scalebox{0.85}{
    \begin{tabular}{ccrrrrrr}
    \toprule
    \multirow{2}[4]{*}{\textbf{Dataset}} & \multirow{2}[4]{*}{\textbf{Split}} & \multicolumn{1}{c}{\multirow{2}[4]{*}{\textbf{\# Images}}} & \multicolumn{5}{c}{\textbf{KL grade}} \\
\cmidrule{4-8}          &       &       & \multicolumn{1}{c}{\textbf{0}} & \multicolumn{1}{c}{\textbf{1}} & \multicolumn{1}{c}{\textbf{2}} & \multicolumn{1}{c}{\textbf{3}} & \multicolumn{1}{c}{\textbf{4}} \\
    \midrule
    \midrule
    OAI   & Train/Val & 39,902 & 15,954 & 7,636 & 9,617 & 5,228 & 1,467 \\
    
    MOST  & Test & 3,445                               & 1,550& 568& 520& 559& 248       \\
\bottomrule
\end{tabular}}

\end{table}

\subsection{Experimental Setup} \label{sc:implementation_details}

\subsubsection{Data Pre-Processing} \label{preprocessing}
\linelabel{ln:preprocessing_in_line_21}\update{We kept our data pre-processing pipeline similar to \cite{tiulpin2018automatic}, and consistently applied it to all the methods in our experiments.}

The essential step in analyzing knee radiographs is a region of interest (ROI) localization. In this paper, we focused rather on the development of an efficient DL architecture and SSL method than on implementing a full knee X-ray analysis pipeline. Thus, we utilized a Random Forest Regression Voting Constrained Local Model approach implemented in the BoneFinder (BF) tool~\cite{lindner2013fully}. \linelabel{ln:bonefinder}\update{For every image, we performed the padding of $300$ pixels on each side to account for position precision in the raw data. After our manual inspection of all the cases, we observed that BF localized all the knees in OAI and MOST except two images (image files could not be read by the software). We show data selection in more detail in \Cref{fig:oai_splitting_workflow}.}


In each bilateral X-ray, we localized the anatomical landmarks using BF and cropped the ROIs of $140mm\times140mm$ centered at each knee joint ($2$ ROIs per image at most). Subsequently, based on the anatomical landmarks, we performed the standardization of each of the ROIs by a horizontal alignment of the tibial plateau.

To standardize the intensity of the histograms, we performed intensity clipping to the $5^{th}$ and $99^{th}$ percentiles as well as global contrast normalization prior to converting $16$-bit raw DICOM images into an 8-bit intensity range. We then center-cropped the obtained $8$-bit images to $110mm\times110 mm$ and resized them to $300\times 300$ pixels ($0.37 mm$ pixel spacing). \linelabel{ln:second_crop}\update{This second cropping aimed to eliminate more background from input images. \linelabel{ln:preprocessing_order}We note that it is also feasible to perform the histogram standardization after the second cropping; however, we followed the order proposed in \cite{tiulpin2018automatic} for better reproducibility of their pipeline, which was retrained in the present study.}

At the next step of the pre-processing, similar to~\cite{tiulpin2018automatic}, we flipped the left images to look like the right ones, and cropped the medial and the lateral patches with a size of $128\times128$ pixels ($H=128$). 
These patches were cropped with the common top anchor at one third of the image height. Both patches were cropped at the lateral and medial image sides. After cropping, we flipped the medial patch horizontally (\cref{fig:architecture}). Finally, we normalized the intensities of each item in the obtained pair into the intensity range of $[-1,1]$. Whereas the previous studies~\cite{tiulpin2018automatic,antony2016quantifying} required the statistics of their training sets to normalize input data, we utilized the mean of $0.5$ and the standard deviation of $0.5$.

\subsubsection{Training and Evaluation Protocol} 
\label{sc:train_model_selection}
\paragraph{Data Split} First, we split the OAI dataset into $2$ parts so that $25\%$ and $75\%$ were for labeled and unlabeled data, respectively. Here, we stratified the splits by KL grade, and ensured that patient IDs do not belong to both of these parts. We then applied another stratification as above to divide each of the aforementioned splits into $5$ folds \linelabel{ln:data_split_edit}\update{(illustrated in \Cref{fig:oai_splitting_workflow})}. In each fold, we generated $24$ data settings, having $4$ labeled ($50$, $100$, $500$, and $1000$ samples per KL grade), and 6 unlabeled data configurations. Here, the amount of unlabeled data was $N$, $2N$, $3N$, $4N$, $5N$, or $6N$ samples, where $N$ is the corresponding amount of labeled data. 

In addition, we also used the whole OAI dataset to assess the performance of our SL baseline. The training and the validation sets of the setting had $31,922$ and $7,980$ samples, respectively. Therefore, the $4$ aforementioned labeled data settings varied from $0.8\%$ to $16\%$ of the whole training set. 

\paragraph{Architecture Selection and Training Setup} Following the protocol proposed by Oliver \textit{et al.}~\cite{oliver2018realistic}, we firstly tuned the SL setting before comparing it to SSL methods. As such, we considered 6 feasible architectures, each of which was the combination of either the architecture of the SL baseline~\cite{tiulpin2018automatic}, or ours with each type of pooling layer (GAP, SAM-VH, or SAM-HV). In our experiments, we selected the top-3 architectures for further comparisons based on cross-validation. The best model among those was utilized as the base model of \textit{Semixup} and SSL baselines. 

\paragraph{SL and SSL Comparisons}
We investigated effects of SSL methods such as the $\Pi$-model~\cite{laine2016temporal}, MixMatch~\cite{berthelot2019mixmatch}, and \textit{Semixup} without the use of unlabeled data. In such scenario, MixMatch~\cite{berthelot2019mixmatch} is equivalent to \textit{mixup}~\cite{zhang2017mixup}. Ultimately, we compared \textit{Semixup} to $3$ SSL baselines such as the $\Pi$ model~\cite{laine2016temporal}, Interpolation Consistency Training (ICT)~\cite{verma2019interpolation}, and MixMatch~\cite{berthelot2019mixmatch}. Each method was trained on the $24$ data settings of the first fold previously generated from the OAI, and finally evaluated on the independent test set from the MOST. 

\paragraph{Metrics} In the training phase of all the experiments, the best models were selected based on both balanced multi-class accuracy (BA)~\cite{urbanowicz2015exstracs} and Cohen's quadratic kappa coefficient (KC)~\cite{cohen1960coefficient}. In the final model evaluation and comparison to the baseline methods, we used BA. To be in line with the metrics used in the previous studies, such as~\cite{antony2018automatic, antony2016quantifying, tiulpin2018automatic}, we also used confusion matrix, KC, and mean square error (MSE). To assess the performance of detecting radiographic OA (KL~$\geq$ 2), we used receiver operating characteristic (ROC) curves, area under the ROC curve (AUC), precision-recall (PR) curves, and average precision (AP). 

\paragraph{Statistical Analyses}

In our initial experiments, we noticed that several factors such as data acquisition center, knee side (left or right), and subject ID may affect the validity of the statistical analyses. To verify this, we used a generalized linear mixed effects model~\cite{breslow1993approximate}, and noticed that neither of these factors has an impact on the results. Therefore, for simplicity of evaluation, we used standard error and one-sided Wilcoxon signed-rank test~\cite{wilcoxon1992individual}. Here, we split the test into $20$ equal-sized chunks (no overlapping patients), and calculated the BA on each of them. Finally, these values were used for the statistical analyses.

\begin{table}[!htbp]
  \centering
  \renewcommand{\arraystretch}{1.2}
  \caption{\small Ablation study of the SAM pooling (SL setting). The values indicate balanced accuracies (\%). The results were \remove{estimated} \update{computed}  out-of-fold using a 5-fold cross-validation. \update{The bold and the underline highlight the best and the second best results, respectively.}}
  \scalebox{0.9}{
  \begin{tabular}{llccccc}
    \toprule
    \multirow{2}{*}{\textbf{Base model}} & \multirow{2}{*}{\textbf{Pooling}} & \multicolumn{5}{c}{\textbf{\# data / KL grade}} \\
\cmidrule{3-7}          &       & \textbf{50} & \textbf{100} & \textbf{500} & \textbf{1000} & \textbf{Average} \\
    \midrule
    \midrule
    \multirow{3}{*}{Tiulpin \etal~\cite{tiulpin2018automatic}} & GAP   &    43.7  & 52.5 &    62.1  &    66.8  &    56.3  \\
     & SAM-VH & \textbf{54.0} &    50.7  &    62.5  & \underline{69.1} & \underline{59.1} \\
     & SAM-HV &    39.0  &    51.2  &    63.7  &    62.7  &    54.1  \\
    \midrule
    \multirow{3}{*}{Ours (Sec. \ref{sc:architecture})}  & GAP   &    46.7  &    49.6  & \textbf{67.0} &    68.6  &    58.0  \\
      & SAM-VH  &    47.7  &    \underline{54.6}  &    65.8  &    65.5  &    58.4  \\
      & SAM-HV & \underline{48.3} & \textbf{56.3} & \underline{66.7} & \textbf{69.7} & \textbf{60.3} \\
    \bottomrule
    \end{tabular}%
}
  \label{tbl:ablation_pooling}%
\end{table}%

\subsection{Supervised Baseline} \label{sc:sl_tuning}
\subsubsection{Model Selection}
\label{sc:sl_pooling_ablation}
\cref{tbl:ablation_pooling} shows that, on average, the models with our base architecture performed $4.1\%$ better compared to the models with~\cite{tiulpin2018automatic} in the case of $500$ samples per KL grade. 
With respect to the base architecture of~\cite{tiulpin2018automatic}, SAM-VH was the most suitable pooling operator, especially in the cases of $50$ and $1000$ samples per KL grade. On the other hand, our base architecture with SAM-HV yielded the highest average BA \linelabel{ln:sam_edit1}\update{across all settings}. 

Based on the average BAs in~\cref{tbl:ablation_pooling}, we selected the three best architectures for further comparisons. The best one among those (ours with SAM-HV) was chosen as the base model of all the SSL methods.

\subsubsection{Performance on the Test Set}

In the first group of results, \cref{tbl:sl_ssl_summary} shows that our architecture with SAM outperformed the SL baseline model~\cite{tiulpin2018automatic} in all the data settings. Specifically, our architecture with SAM-HV yielded BAs $9\%$ better than the baseline~\cite{tiulpin2018automatic} in the data settings of $500$ and $1000$ images per KL grade. Notably, in the case of $500$ samples per KL grade, our model with SAM-HV surpassed by $6\%$ the baseline model that would require a double amount of data. Finally, our model with SAM-HV achieved a BA of $67.5\%$ when trained on only $16\%$ of the full OAI set. 

\begin{table}[htbp]
  \centering
  \renewcommand{\arraystretch}{1.3}
  \caption{\small  SL and SSL methods evaluation on the test set.
  We reported the top-3 SL models with GAP$^{\star}$, SAM-VH$^{\dagger}$, or SAM-HV$^{\ddagger}$. Our model with SAM-HV$^{\ddagger}$ is the common architecture for all SSL methods. Bold highlights models performing substantially better than the second best model in each category. $\diamond$ indicates no substantial difference among the best models.}
    \scalebox{0.9}{
    \begin{tabular}{lllll}
    \toprule
    \multicolumn{1}{l}{\multirow{2}{*}{\textbf{Method}}} & \multicolumn{4}{c}{\textbf{\# labels / KL grade}} \\
    \cmidrule{2-5}
          & \multicolumn{1}{c}{\textbf{50}} & \multicolumn{1}{c}{\textbf{100}} & \multicolumn{1}{c}{\textbf{500}} & \multicolumn{1}{c}{\textbf{1000}} \\
    \midrule \midrule
    \multicolumn{5}{l}{Fully SL} \\
    \midrule
    \multicolumn{1}{l}{Tiulpin \etal$^{\star}$~\cite{tiulpin2018automatic}} & 40.5$\pm$0.8 & 49.7$\pm$0.9 & 55.1$\pm$0.8 & 58.5$\pm$0.8 \\          
    \multicolumn{1}{l}{Tiulpin \etal$^{\dagger}$~\cite{tiulpin2018automatic}} & \secondbest{45.2$\pm$0.8} & 48.6$\pm$0.9 & 57.6$\pm$0.8 & 58.5$\pm$0.8 \\          
    \multicolumn{1}{l}{Our SL$^{\dagger}$ (Sec. \ref{sc:architecture})} & \secondbest{45.6$\pm$0.8} & \secondbest{53.2$\pm$0.9} & 61.5$\pm$0.8 & 63.5$\pm$0.8 \\          
    \multicolumn{1}{l}{Our SL$^{\ddagger}$ (Sec. \ref{sc:architecture})} & 41.5$\pm$0.8 & \secondbest{52.9$\pm$0.9} & \textbf{64.0$\pm$0.8} & \textbf{67.5$\pm$0.8} \\          
    \midrule \midrule
    \multicolumn{5}{l}{SL and SSL methods - Without unlabeled data} \\
    \midrule
    \multicolumn{1}{l}{\textit{mixup}~\cite{zhang2017mixup}} & 39.6$\pm$0.8 & 53.9$\pm$0.8 & 64.5$\pm$0.8 & \secondbest{67.1$\pm$0.8}  \\          
    \multicolumn{1}{l}{$\Pi$ model~\cite{laine2016temporal}} & \textbf{42.3$\pm$0.8} & \secondbest{56.3$\pm$0.8} & 64.3$\pm$0.8 & \secondbest{67.9$\pm$0.8} \\     
    \multicolumn{1}{l}{\textit{Semixup} (Ours)} & 31.7$\pm$0.8 & \secondbest{56.1$\pm$0.8} & \textbf{66.6$\pm$0.8} & \secondbest{68.0$\pm$0.8} \\          
    \midrule \midrule
    \multicolumn{5}{l}{SSL methods - Best performing models' comparison} \\
    \midrule
    \multicolumn{1}{l}{ICT~\cite{verma2019interpolation}} & \secondbest{47.7$\pm$0.9} & 53.5$\pm$0.8 & 64.1$\pm$0.8 & 67.8$\pm$0.8 \\          
    \multicolumn{1}{l}{$\Pi$ model~\cite{laine2016temporal}} & \secondbest{45.9$\pm$0.8} & 56.2$\pm$0.8 & 67.1$\pm$0.8 & 69.0$\pm$0.8 \\          
    \multicolumn{1}{l}{MixMatch~\cite{berthelot2019mixmatch}} & \secondbest{45.1$\pm$0.8} & 56.0$\pm$0.8  & 67.6$\pm$0.8 & 68.4$\pm$0.8 \\          
    \multicolumn{1}{l}{\textit{Semixup} (Ours)} & \secondbest{46.9$\pm$0.9} & \textbf{58.8$\pm$0.8} & \textbf{69.7$\pm$0.8} & \textbf{71.0$\pm$0.8} \\         
	\midrule \midrule
    \multicolumn{5}{l}{Fully SL - Trained on the full OAI (31,922 samples)} \\
    \hline
    Our SL$^{\ddagger}$ (Sec. \ref{sc:architecture})& \multicolumn{4}{c}{70.9$\pm$0.8} \\
    \bottomrule
    \end{tabular}}
  \label{tbl:sl_ssl_summary}%
\end{table}%

\subsection{Semi-Supervised Learning} \label{sc:compare_sl_ssl}

\linelabel{ln:remove_not_relavence}\remove{We summarized the results of SL and SSL models trained on 4 labeled data settings, and the SL model trained on full OAI data in Table III. Detailed performance evaluations of all the SSL methods are described in Table S8. }

\subsubsection{SSL Methods without Unlabeled Data}

\linelabel{ln:ssl_0x_update1}\update{We summarized our results in \cref{tbl:sl_ssl_summary}.} The performance of \textit{Semixup} scaled well with the amount of training data, and outperformed its corresponding fully SL model when we had at least $100$ samples per KL grade. 
In the comparison with other SSL baselines, \textit{Semixup} achieved comparable results \linelabel{ln:ssl_0x_update2}\update{when having no unlabeled data at all.}

\subsubsection{Semixup Compared to SL and SSL Baselines}
\label{sc:semixup_vs_sl_ssl}

After evaluating $24$ trained models of each SSL method, we derived maximum BAs grouped by labeled data setting as in the third part of \cref{tbl:sl_ssl_summary}. With $500$ labeled samples per KL grade, \textit{Semixup} achieved a BA of $69.7\%$, which was $5.7\%$ higher than the result of the SL baseline, and surpassed the SL baseline by $2.2\%$ even when trained on $2$ times less labeled data than the baseline was. With $1000$ labeled \linelabel{ln:sl_ssl_typo1}\remove{data}\update{samples} per KL grade, our SSL method reached the BA of $71\%$ that was comparable to the result of the SL baseline trained on the full OAI setting (over $6$ times more data used).

\begin{figure}[h]
    \centering
    \IfFileExists{figs/comparison_sl_ssl_ci-crop.pdf}{}{\immediate\write18{pdfcrop figs/comparison_sl_ssl_ci.pdf}}
        \includegraphics[scale=0.85]{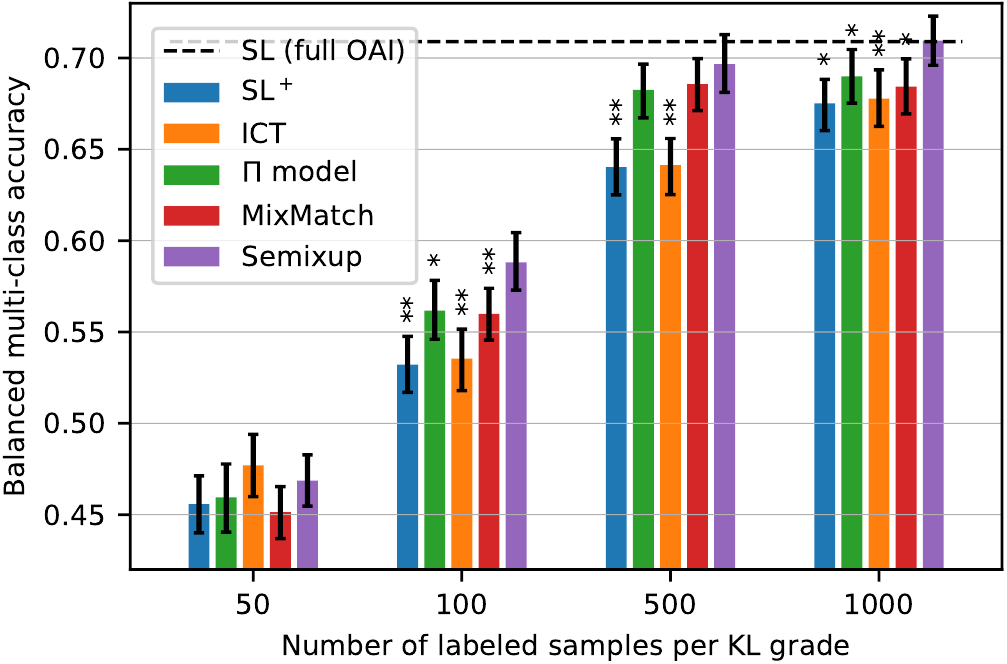}
    \caption{\small Graphical comparison of \textit{Semixup} and the baseline methods (MOST dataset). The bars indicate the 95\% confidence intervals, SL$^{+}$ indicates our fully SL models equipped with either SAM-HV or SAM-VH, and * and ** indicate the statistically significant difference ($p<0.05$ and $p<0.001$, respectively). The dash line indicates the BA of the fully SL model with SAM-HV trained on the full OAI dataset.}
    \label{fig:best_sl_ssl_comparison_ci}
\end{figure}

\begin{figure}[h]
    \centering
    \IfFileExists{figs/cm/CM_semixup_500_1500-crop.pdf}{}{\immediate\write18{pdfcrop figs/cm/CM_semixup_500_1500.pdf}}
    \IfFileExists{figs/cm/CM_semixup_1000_1000-crop.pdf}{}{\immediate\write18{pdfcrop figs/cm/CM_semixup_1000_1000.pdf}}
       
        \subfloat[500 / KL grade
        \label{fig:cm500}]{\includegraphics[scale=0.21]{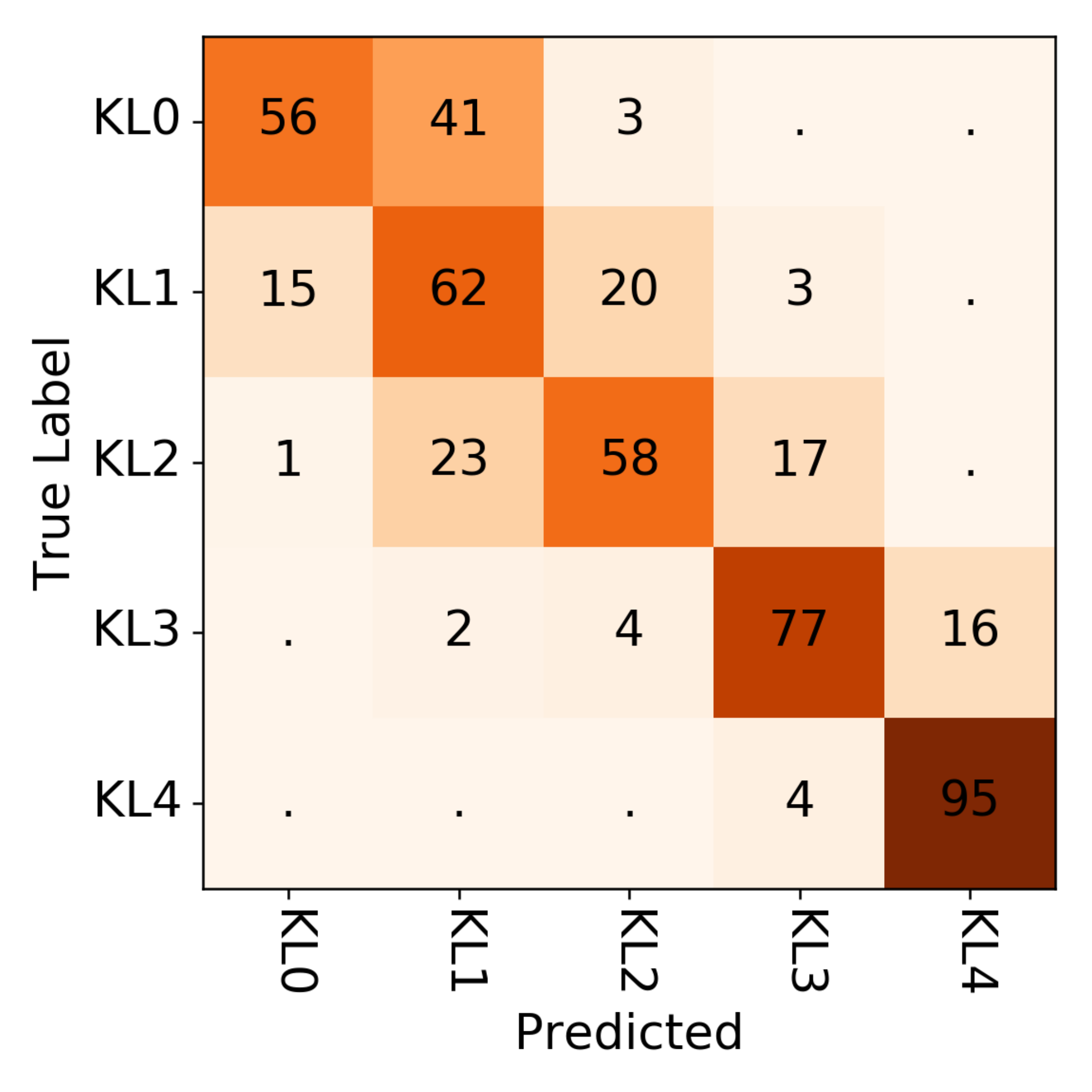}}\hfill%
        \subfloat[1000 / KL grade
        \label{fig:cm1000}]{\includegraphics[scale=0.21]{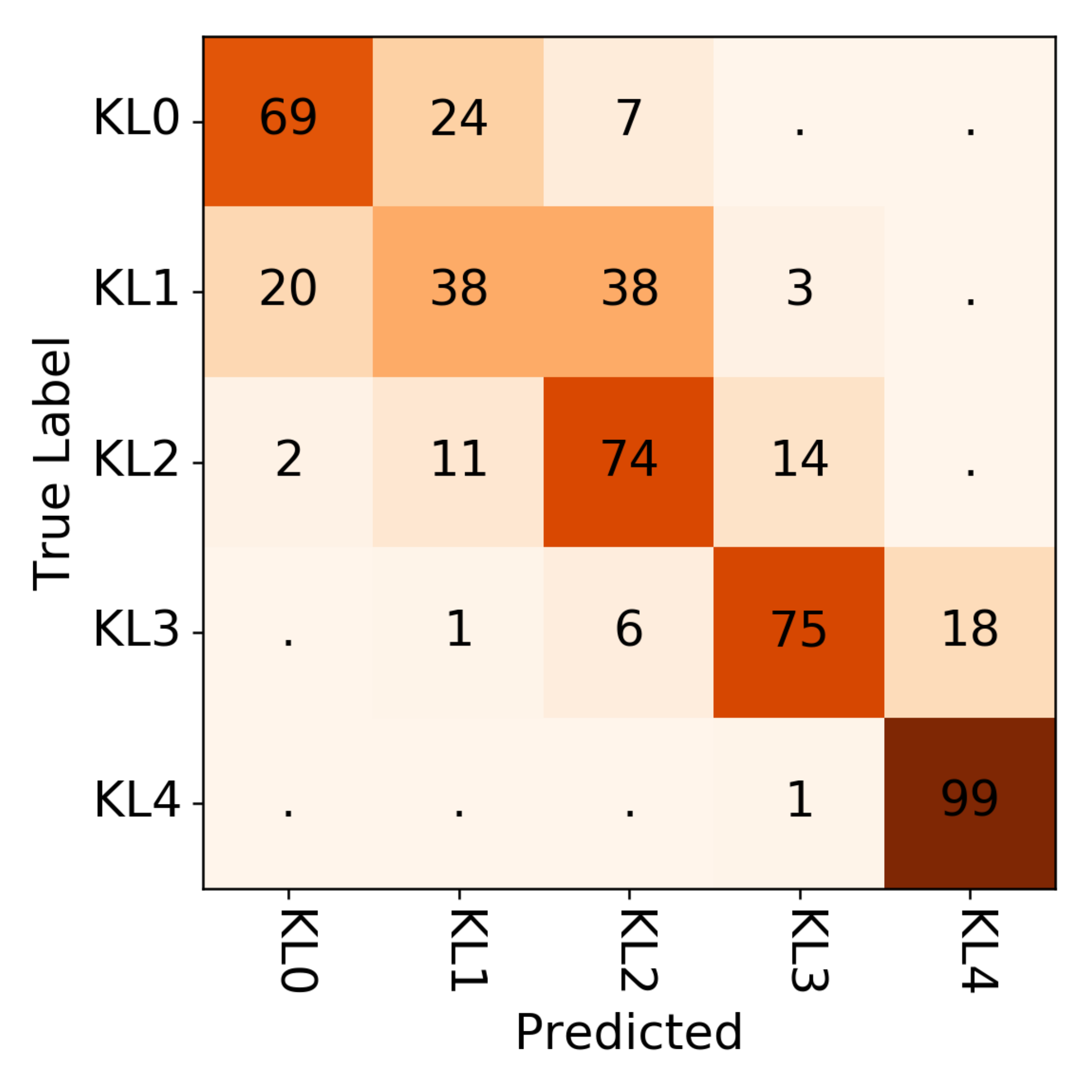}}
        \hspace*{\fill}%
    \caption{\small Confusion matrices showing performance (\%) of our models trained with \textit{Semixup} on the test set (MOST dataset) in two labeled data settings (\protect\subref*{fig:cm500}) and (\protect\subref*{fig:cm1000}).}
    \label{fig:best_semixup_cm}
\end{figure}

\begin{table}[h]
  \centering
  \caption{Comparison of our best models trained with \textit{Semixup} against our well-tuned SL model with SAM-HV. }
  \scalebox{0.73}{
    \begin{tabular}{lrcccc}
    \toprule
    \multicolumn{1}{c}{\textbf{Method}} & \multicolumn{1}{c}{\textbf{\# labels}} & \multicolumn{1}{c}{\textbf{KC}} & \multicolumn{1}{c}{\textbf{MSE}} & \multicolumn{1}{c}{\textbf{AUC (KL $\geq$ 2)}} & \multicolumn{1}{c}{\textbf{AP (KL $\geq$ 2)}} \\
    \midrule
    \midrule
    \multicolumn{1}{c}{\multirow{8}{*}{Semixup}} & \multirow{2}{*}{250}   & 0.708  & 1.080  &    0.880    &    0.861    \\
    &   &  [0.692, 0.724] & [1.000, 1.022] & [0.869, 0.891] & [0.849, 0.872] \\
          & \multirow{2}{*}{500}   & 0.789 & 0.810  &    0.933    &    0.916    \\
    &   & [0.776, 0.801] & [0.764, 0.860] & [0.926, 0.940] & [0.906, 0.925] \\
          & \multirow{2}{*}{2,500}  & 0.877 & 0.442  &    0.967    &    0.956    \\
    &   & [0.870, 0.884] & [0.416, 0.466] & [0.962, 0.972] & [0.950, 0.962] \\
          & \multirow{2}{*}{5,000}  & 0.878  & 0.458  &    0.972   &    0.959   \\
    &   & [0.870, 0.885] & [0.430, 0.487] & [0.968, 0.976]  & [0.953, 0.965] \\
    \midrule
    \multicolumn{1}{c}{\multirow{2}{*}{SL (full OAI)}} & \multirow{2}{*}{31,922} & 0.881  & 0.440  &    0.974    &    0.963    \\
    &   & [0.873, 0.889] & [0.414, 0.471] & [0.969, 0.978] & [0.958, 0.969] \\
    \bottomrule
    \end{tabular}}%
  \label{tbl:comparison_many_metrics}%
\end{table}%

Besides comparing to the SL settings, we also compared \textit{Semixup} to the state-of-the-art SSL baselines. On the test set, our method outperformed the others in the data settings of $100$, $500$, and $1000$ labels per KL grade. In the data setting of $2500$ labeled images ($7.8\%$ of the full OAI size), \textit{Semixup} yielded a BA 2.1\% higher than the best SSL baseline, MixMatch, with the same labeled data amount, and $0.7\%$ higher than the best SSL baseline, the $\Pi$ model, with twice the amount of labeled data. In addition, \textit{Semixup} achieved the highest average BAs in the data settings having at least $100$ labeled images per KL grade (\Cref{tbl:detailed_comparison}).

The comparisons between \textit{Semixup} and each of the baseline models across different labeled data settings are graphically illustrated in \cref{fig:best_sl_ssl_comparison_ci}. The best models of \textit{Semixup} in the cases of $100$ and $1000$ labels per KL grade  significantly outperformed all the baselines ($p < 0.05$). In the case of $500$ labeled samples per KL grade, \textit{Semixup} was significantly better than the SL baseline and ICT with the same amount of labels. Furthermore, our method also surpassed the SL baseline that was trained with twice more labeled data ($1000$ samples per KL grade). The statistical analyses indicate that SL model trained on the full OAI dataset did not differ significantly from the \textit{Semixup} models that were trained with  $500$ and $1000$ labeled samples per KL grade ($p$-values were of $0.054$ and $0.368$, respectively). The details of the statistical testing are presented in \Cref{subtbl:detailed_wilcoxon_test}.

\linelabel{ln:more_data_note}\update{We finally note here that we also evaluated larger amounts of labeled data. However, we observed a saturation of performance increase with their further addition to the training. Therefore, we omitted these results from the manuscript, but they can be found in Supplement (\Cref{sc:more_labeled_data})}.

\subsubsection{Detection of Early Radiographic Osteoarthritis}\label{sc:semixup_more}
\cref{fig:best_semixup_cm} presents the confusion matrices of our 2 best models. Here, we evaluate the accuracy of our method to detect early OA (i.e., KL~$=2$). With $500$ and $1000$ labels per KL grade, \textit{Semixup} was able to detect early OA with $58\%$ and $74\%$ accuracies. Notably, without the doubtful OA (KL~$=1$), we achieved a substantially high BA of $79.25\%$.

\begin{figure}[htbp]
    \centering
    \IfFileExists{figs/roc_auc/ROC_AUC_Semixup-crop.pdf}{}{\immediate\write18{pdfcrop figs/roc_auc/ROC_AUC_Semixup.pdf}}
    \IfFileExists{figs/pre_rec/PRE_REC_Semixup-crop.pdf}{}{\immediate\write18{pdfcrop figs/pre_rec/PRE_REC_Semixup.pdf}}
        
        \subfloat[ROC curve
        \label{fig:roc_auc}]{\includegraphics[scale=0.71]{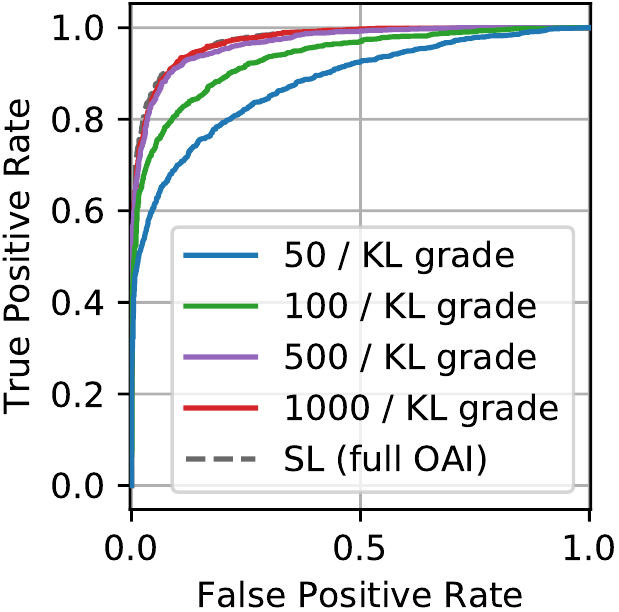}} 
    \subfloat[PR curve
        \label{fig:pre_rec}]{\includegraphics[scale=0.71]{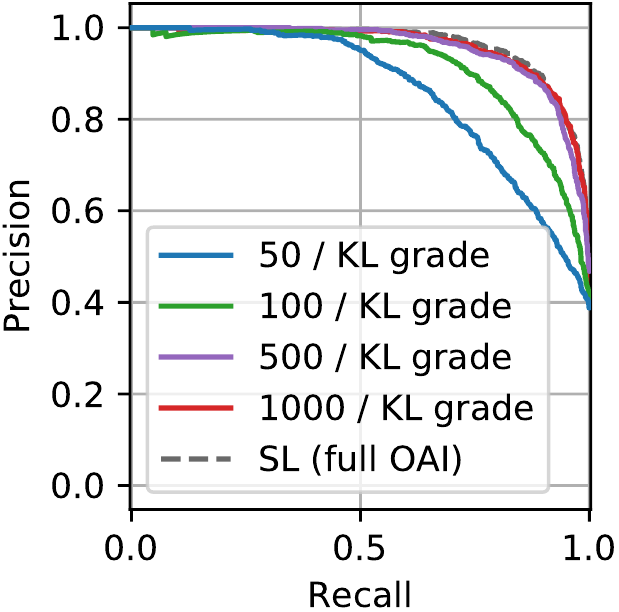}}
    \caption{\small Comparison of the best models of trained with \textit{Semixup} on radiographic OA detection task (KL $\geq$ 2). The models were trained with 50, 100, 500, or 1000 labeled examples per KL grade.}
    \label{fig:semixup_roc_auc}
\end{figure}

 \linelabel{ln:data_efficient_2}With regard to detection of radiographic OA (KL $\geq2$),~\cref{fig:semixup_roc_auc} and \cref{tbl:comparison_many_metrics} show how our SSL model with $500$ labels per KL grade was comparable to the well-tuned SL model trained on the large training set (more than $12$ times labeled samples). The detailed comparisons with respect to different amounts of unlabeled data are presented in \Cref{supfig:semixup_roc_auc} and \Cref{supfig:semixup_pre_rec}.

\subsection{Ablation Study}
\label{sc:ablation_study}
\subsubsection{Impact of Consistency Regularization Terms}\label{sc:ablation_clustering}

Because our in- and out-of-manifold regularizer comprises several individual losses, it is essential to understand the impact of each of them onto the method's performance. To assess the contributions of each used regularizer -- \eqref{eqn:consistency},~\eqref{eqn:ict}, and~\eqref{eqn:semixup_term}, we consecutively removed each of them from our loss function. \cref{tbl:ablation_loss} shows this ablation study. Here, we report the best validation performance among 6 different unlabeled data settings. The results show that \textit{Semixup} performs better when all the regularizers are used jointly.

\begin{table}[htbp]
\renewcommand{\arraystretch}{1.2}
  \centering
  \caption{\small Ablation study of regularization terms used in \textit{Semixup}. The values are BAs (\%).}
  \begin{tabular}{lcc}
    \toprule
    \multirow{2}{*}{\textbf{Method}} & \multicolumn{2}{c}{\textbf{\# data / KL grade}} \\
\cmidrule{2-3}          & \multicolumn{1}{c}{\textbf{100}} & \multicolumn{1}{c}{\textbf{500}} \\
    \midrule
    \midrule
    Semixup w/o in-manifold terms &        60.8  &        73.1 \\
    Semixup w/o out-of-manifold term &        60.8  &        72.0  \\
    Semixup w/o interpolation consistency &        60.8  &        73.4  \\
    Semixup & \textbf{65.6} & \textbf{74.1} \\
    \bottomrule
    \end{tabular}%
  \label{tbl:ablation_loss}%
\end{table}

\subsubsection{Balancing of Regularization Coefficients}
\label{sc:reg_coef_semixup}

\update{\cref{tbl:ablation_loss} shows that all the  consistency regularization terms used in our method in are needed to reach good performance. However, it is unclear what practical effect the balancing of these terms may have on the performance of our method. To gain insights into what value ranges of these coefficients are appropriate, we performed additional  sensitivity analyses of these hyperparameters by varying the values used in our experiments.}

\begin{figure}[ht]
    \centering
    \IfFileExists{figs/cm/grid_search-crop.pdf}{}{\immediate\write18{pdfcrop figs/cm/grid_search.pdf}}
        \includegraphics[scale=0.50]{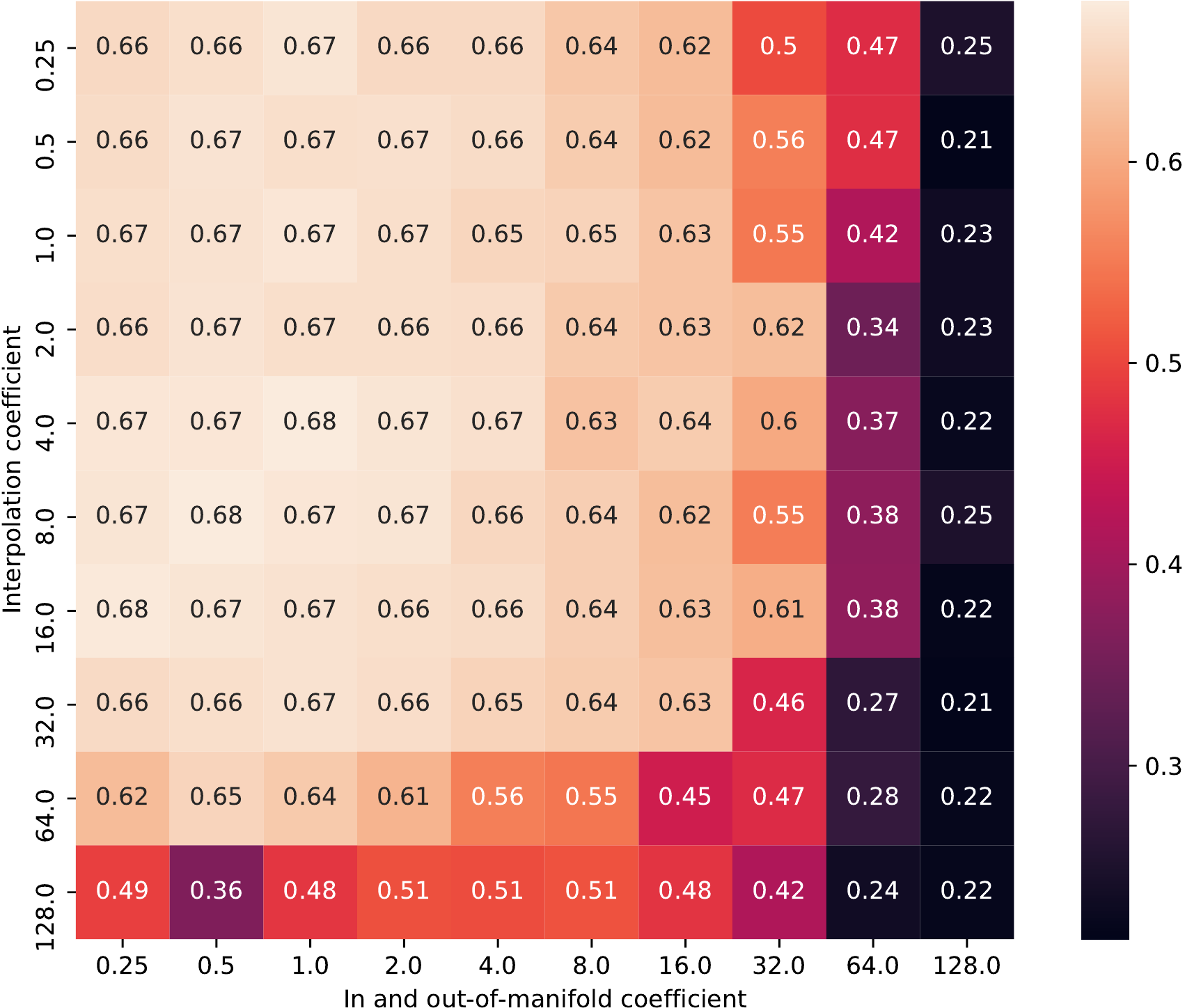}
    \caption{\small \update{Sensitivity analysis of balancing interpolation consistency term's coefficient versus the in- and out-of-manifold ones. We report the average balanced accuracy on validation set over 5 random seeds for the case of $2500$ labeled, and $2500$ unlabeled samples. The weights of in- and out-of-manifold terms were set to the same value.}}
    \label{fig:weight_grid_search}
\end{figure}

\begin{figure}[ht]
    \centering
    \IfFileExists{figs/grid_search_iom-crop.pdf}{}{\immediate\write18{pdfcrop figs/grid_search_iom.pdf}}
        \includegraphics[scale=.6]{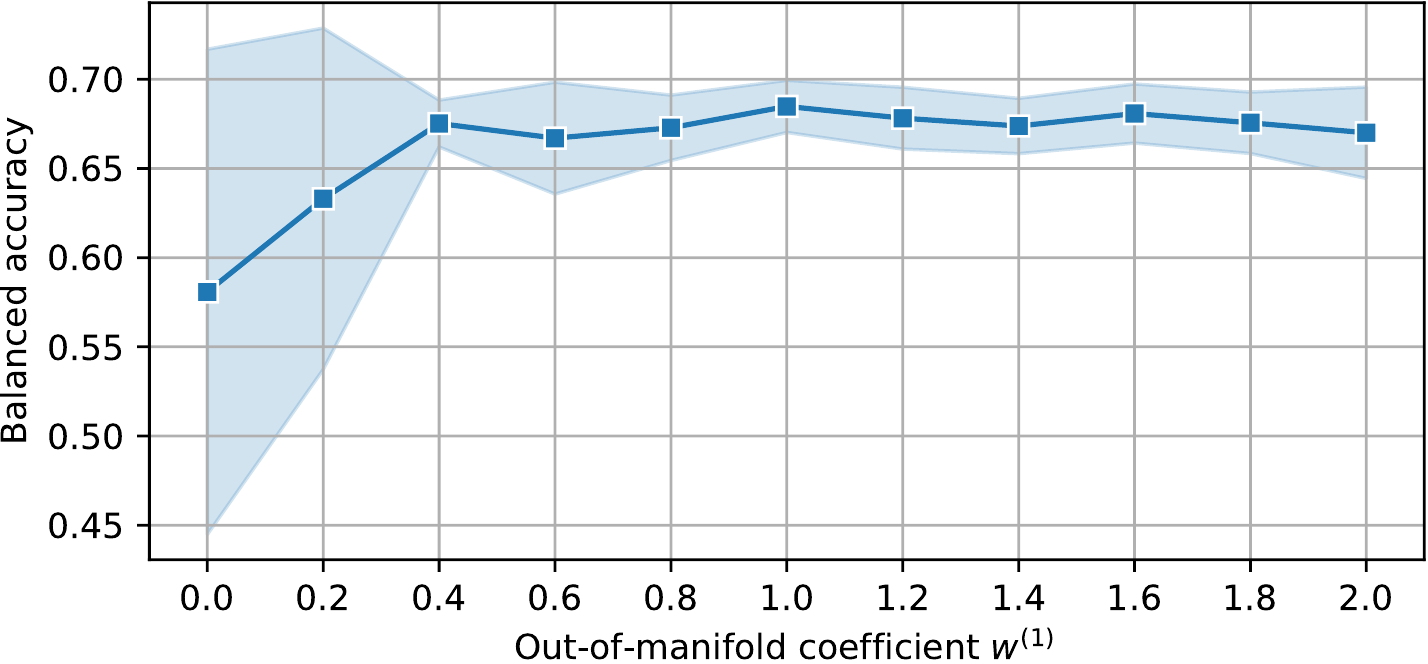}
    \caption{\small \update{Relationship between in- and out-of manifold terms. Interpolation consistency term's coefficient was set to $4$, and the sum of in- and out-of-manifold consistency terms' weights was set to $2$. We evaluated the average balanced accuracy on the validation set over 5 different random seeds for $2500$ labeled, and $2500$ unlabeled samples. The solid line and the band indicate the mean and standard deviation, respectively.}}
    \label{fig:weight_grid_search_iom}
\end{figure}

\update{Firstly, we showed the trade-off between the interpolation consistency term and the rest (\cref{fig:weight_grid_search}). Secondly, we showed the effect of balancing in- and out-of-manifold consistency regularization terms (\cref{fig:weight_grid_search_iom}). All the experiments in this subsection demonstrate the performance on the validation set in the case of $2500$ labeled samples ($500$ per KL-grade), and the same amount of unlabeled samples.}


\section{Discussion and Conclusions}
\label{sc:dis_conc}
In this study, we presented a novel SSL method -- \textit{Semixup}. This method leverages in- and out-of-manifold consistency regularization, and we demonstrated its application in the task of automatic grading of knee OA severity. Furthermore, we also proposed a novel state-of-the-art architecture for this task.

The core novelty of this work lies in using \textit{mixup} for generating out-of-manifold samples in close proximity of data manifold, and then ensuring consistent predictions made by a neural network on such samples and the ones drawn from the data manifold itself. We experimentally showed that the proposed technique helps to train more robust models in limited data regimes even when unlabeled data is not available. 

\linelabel{ln:data_efficient_1}\update{To be specific about the label efficiency and robustness, we have shown that the \textit{Semixup}-trained model using $5000$ labeled  samples ($1000$ per KL-grade) performs comparably to the well-tuned supervised baseline trained on the full OAI dataset (\Cref{subtbl:detailed_wilcoxon_test}, $p=0.368$). Furthermore, \cref{tbl:sl_ssl_summary} and \cref{tbl:comparison_many_metrics} show that our SSL model has overlapping confidence intervals with the SL model according to various metrics, which also supports this statement. }

Another novelty of this work is the proposed pooling approach that allows to efficiently process large feature maps, thereby leveraging fine-grained information required for knee OA grading. Our work demonstrated not only the new method for SSL, but also an efficient baseline for supervised setting when the amount of training samples is limited. To our knowledge, this is the first work that systematically studied the problem of learning robust classifiers for knee OA severity assessment in the limited data regime.  

\linelabel{ln:apply_semixup_to_others}\update{We note that while we leveraged the domain knowledge to develop a base CNN architecture (Siamese network), our SSL method is not constrained to any specific domain. Thus, we think that \textit{Semixup} can be used with other data (e.g., in CT or MRI related tasks). However, we also highlight that this requires an experimental evidence, producing which is beyond the scope of our study. To facilitate the applications of our method in other fields, we release the source code at \url{https://github.com/MIPT-Oulu/semixup}.} \remove{Our implementation is publicly available at \url{https://github.com/MIPT-Oulu/semixup}.}

\linelabel{ln:discussion_methodological_impact}\update{From the methodological point of view, our work has several important insights. Firstly, the results shown in~\Cref{suptab:results-all} indicate that none of the evaluated SSL methods consistently scale in performance with scaling the amounts of unlabeled data. Thus, we suggest that it is sufficient to use as many unlabeled samples. The experiments shown in this work (\cref{fig:weight_grid_search_iom}) show that the out-of-manifold consistency regularization is rather more important than the in-manifold one. Here, we point to the recent work by
~Mao~\etal~\cite{mao2019virtual} in the realm of domain adaptation, which showed that using consistency regularization of mixed-up samples imposes a locally-Lipschitz constraint to the model in the regions of close proximity to the data manifold. To conclude, we believe that further exploiting this type of consistency regularization can help in the other domains.}

Besides all the aforementioned benefits, this study has several limitations. Specifically, we did not use EMA in our methods as it is costly to train and the methods using EMA would require significantly more computational power than their competitors. Another limitation of this study is that we did not leverage the power of SAM-VH and SAM-HV in our model. We foresee potential improvements of our results if both of these pooling schemes are \remove{be }used, e.g. by ensembling the results produced by models trained with SAM-VH and SAM-HV, respectively. 

\linelabel{ln:dis_ssl_limitation}\update{A general criticism also applies to our work. While considering SSL to be a powerful tool, we also confirm the concerns previously raised Oliver~\etal~\cite{oliver2018realistic}. Specifically, we emphasize that a thorough evaluation of SSL vs SL should always be done, and one might always reach the best possible performance when using the former one. Therefore, we think that a further research on the stability of SSL methods is needed. However, our results show that SSL was still outperforming SL, especially in the low-data regime.}

From a clinical point of view, using \linelabel{ln:dis_wording_1}\update{composite} KL grades as a reference can also be considered a limitation, thus we suggest future studies to focus on OARSI grading of the knee joints~\cite{altman2007atlas, tiulpin2019automatic}. Finally, another clinical limitation of this work is that OA is typically graded at a late stage, when it is already present, and we think that the next steps should rather focus on developing models for predicting OA progression~\cite{tiulpin2019multimodal} using SSL and partially labeled data. These data are available in hospital archives and can be leveraged at a low cost.

To conclude, we would like to highlight 
the clinical implications of our work. 
Firstly, we demonstrated that highly accurate KL grading can be done with only small amounts of labeled data, which allows small research teams and medical device vendors to build generalizable models suitable for clinical use. \linelabel{ln:dis_remove_1}\remove{We highlight that this work demonstrates results drastically outperforming all the previous studies on automatic KL grading, and also human-level agreement (KC of $0.56-0.85$~\cite{klara2016reliability}), while the proposed method requires drastically less data than has been previously used.} Secondly\update{,} 
the proposed method can significantly reduce 
routine work done by radiologists, which on the societal level can lead to cost savings while at the same time improving the quality of health care. Thirdly, we expect the proposed method and the network architecture to generalize to other domains, such as \linelabel{ln:dis_wording_2}\remove{hips and hands}\update{hip and hand OA}. Finally, we think that this study is an important step towards data efficient medical image recognition, which is currently lacking thoroughly validated methodologies.

\section*{Acknowledgments}
The OAI is a public-private partnership comprised of five contracts (N01- AR-2-2258; N01-AR-2-2259; N01-AR-2- 2260; N01-AR-2-2261; N01-AR-2-2262) funded by the National Institutes of Health, a branch of the Department of Health and Human Services, and conducted by the OAI Study Investigators. Private funding partners include Merck Research Laboratories; Novartis Pharmaceuticals Corporation, GlaxoSmithKline; and Pfizer, Inc. Private sector funding for the OAI is managed by the Foundation for the National Institutes of Health. 

The MOST is comprised of four cooperative grants (Felson - AG18820; Torner - AG18832; Lewis - AG18947; and Nevitt - AG19069) funded by the National Institutes of Health, a branch of the Department of Health and Human Services, and conducted by MOST study investigators. This manuscript was prepared using MOST data and does not necessarily reflect the opinions or views of MOST investigators. 

We would like to acknowledge the strategic funding of the University of Oulu, KAUTE foundation and Sigrid Juselius Foundation, Finland.

Dr. Claudia Lindner is acknowledged for providing BoneFinder. Iaroslav Melekhov is acknowledged for proposing the ablation study of the pooling method. Phuoc Dat Nguyen is acknowledged for discussions about \textit{mixup}.


\bibliography{bibtex/IEEEabrv.bib,bibtex/reference.bib}{}
\bibliographystyle{IEEEtran}

\newpage
\setcounter{page}{1}
\setcounter{figure}{0}
\setcounter{table}{0}

\renewcommand{\thepage}{S\arabic{page}} 
\renewcommand{\thesection}{S\arabic{section}}  
\renewcommand{\thetable}{S\arabic{table}}  
\renewcommand{\thefigure}{S\arabic{figure}}

\section*{Supplement} 
\label{sc:supplement}

\subsection{Experimental Details}\label{ssec:details}
All our experiments were conducted on V100 NVidia GPUs. We used PyTorch, Collagen~\cite{collagen2019}, and SOLT~\cite{tiulpin2019solt} libraries in our codebase. To train all the models, we utilized the Adam optimizer~\cite{kingma2014adam} with a learning rate of $1e\text{-}4$ and without weight decay regularization. Except for the SL baseline model~\cite{tiulpin2018automatic}, which had a dropout of $0.2$ in the bottleneck of the model, we set the dropout rate to $0.35$, and used it in multiple blocks our model. Note that we did not use Exponential \linelabel{ln:sup_typo_1}\update{Moving} Averaging (EMA)~\cite{tarvainen2017mean} in any methods. For \textit{Semixup}, we sampled $1$ pair of random augmentations (i.e., $N_t=2$) and $1$ mixing operator (i.e., $N_m=1$) for each knee image. We set the parameter $\alpha$ to $0.75$, and the unsupervised weight vector $\mathbf{w}$ to $[2, 2, 4]^\top$ (Supplementary Algorithm~\ref{alg:semixup}).

In the pre-processing step, we transformed images by random Gaussian noise, rotation, cropping, and Gamma correction augmentations, whose parameters are described in \Cref{tbl:augmentation}. A description of our network architecture is described in \Cref{tbl:detailed_arch}. Detailed implementations of related methods are expressed in \Cref{sc:implementation_baseline_methods}.

\subsection{Baseline Methods} \label{sc:implementation_baseline_methods}
We reimplemented the previous SSL methods (\url{https://github.com/MIPT-Oulu/semixup}) in a common codebase as suggested by~\cite{oliver2018realistic}. All the baseline methods used the same architecture and pre-processing operators as our approach. We trained each method with a batch size of $40$ for $500$ epochs, each of which is a full pass through labeled data. 

The weight of consistency regularization terms, $\mathbf{w}$, varied among the methods. As such, for $\Pi$-model, we searched it through the set $\{1, 10, 50\}$, and found that it worked well with $w=1$. In ICT, we varied $w$ from $0$ to $100$ using a sigmoid ramp-up schedule within the first $80$ epochs as in~\cite{verma2019interpolation}. Following Verma~\etal, we did not use \textit{mixup} for labeled data in ICT. For MixMatch, we used $2$ augmentations, a sharpening hyperparameter of $0.5$ ($T$ in the original paper), and a $\textrm{Beta}(\alpha, \alpha)$ distribution with $\alpha$ of $0.75$. We run a search for its unsupervised coefficient from $\{1, 10, 50\}$ based on~\cite{berthelot2019mixmatch}, and found the coefficient of $10$ to be the best choice. In addition, we trained \textit{mixup} using a $\textrm{Beta}(\alpha, \alpha)$ distribution with $\alpha$ of $0.75$.

\begin{table}[htbp]
  \centering
  \caption{Ordered list of augmentations}
    \begin{tabular}{clcr}
    \toprule
    \textbf{Order} & \multicolumn{1}{c}{\textbf{Augmentation}} & \textbf{Probability} & \multicolumn{1}{c}{\textbf{Parameter}} \\
    \midrule
    1     & Gaussian noise & 0.5   & 0.3 \\
    2     & Rotation & 1     & [-10, 10] \\
    3     & Padding & 1     & 5\% \\
    4     & Cropping  & 1     & 128$\times$128 \\
    5     & Gamma correction & 0.5   & [0.5, 1.5] \\
    \bottomrule
    \end{tabular}%
  \label{tbl:augmentation}%
\end{table}

\begin{algorithm}[htbp]
\renewcommand{\arraystretch}{1.5}
\SetAlgoNoLine
\KwData{$\mathcal{X}_{ul}$: combined labeled and unlabeled examples}
\KwData{$(\mathcal{X}_l, \mathcal{Y})$: labeled examples}
\KwInput{$f_\theta$: a neural network with parameters $\theta$.}
\KwInput{$\alpha$: parameter of Beta distribution}
\KwInput{$\mathbf{w}$: weights of consistency terms}
\KwInput{$N_T$: the number of iterations}
\KwInput{$N_c$: the number of classes}
\KwInput{$N_b$: batch size}
\For{$t=1\dots N_T$}{
$B_l \xleftarrow{} \{(x_i, y_i)\}_{i=1}^{N_b} \subset (\mathcal{X}_l, \mathcal{Y})$, $B_{ul} \xleftarrow{} \{x_i\}_{i=1}^{N_b} \subset \mathcal{X}_{ul}$\\ 
    $\mathcal{L}_{l}, \mathcal{L}_{u},  \xleftarrow{} 0, 0$ \\
    \For{$x_i \in B_{ul}$}    
    { 
        Sample $x_j \in B_{ul}$, $\lambda \sim \textrm{Beta}(\alpha, \alpha)$, $T$ and $T' \sim p(\tau)$ \\
        $\lambda \xleftarrow{} \max(\lambda, 1 - \lambda)$ \\
        $x_{mix}\xleftarrow{} \textrm{Mix}_{\lambda}(Tx_i, x_j)$ \\
        $p_{mix} \xleftarrow{} \textrm{Mix}_\lambda(f_{\theta}(Tx_{i}), f_{\theta}(x_{j})) $ \\
    	$\mathcal{L}_u \xleftarrow{} \mathcal{L}_u + \mathbf{w}^{(0)}\left \| f_{\theta}(Tx_i) - f_{\theta}(T'x_i) \right \|_{2}^{2}$\\

    	$\mathcal{L}_u \xleftarrow{} \mathcal{L}_u + \mathbf{w}^{(1)}\left \| f_{\theta}(x_{mix}) - f_{\theta}(Tx_i)\right \|_{2}^{2}$\label{algo_line:semixup_1} \\
    	
    	$\mathcal{L}_u \xleftarrow{} \mathcal{L}_u + \mathbf{w}^{(1)}\left \| f_{\theta}(x_{mix}) - f_{\theta}(T'x_i)\right \|_{2}^{2}$\label{algo_line:semixup_2} \\

    	$\mathcal{L}_u \xleftarrow{} \mathcal{L}_u + \mathbf{w}^{(2)}\left \| f_{\theta}(x_{mix}) - p_{mix} \right \|_{2}^{2}$\\
    }
    
    \For{$(x_i, y_i)  \in B_l$}    
    { 
        Sample $(x_j, y_j) \in B_{l}$, $\lambda \sim \textrm{Beta}(\alpha, \alpha)$\\
        $x_{mix} \xleftarrow{} \textrm{Mix}_{\lambda}(x_i, x_j)$ \\
    	$\mathcal{L}_l \xleftarrow{} \lambda \mathcal{L}_{ce}\left(x_{mix}, y_i\right) + (1 - \lambda) \mathcal{L}_{ce}(x_{mix}, y_j)$ \\
    }
    $\mathcal{L} \xleftarrow{} \frac{1}{N_{b}}\mathcal{L}_l + \frac{1}{N_{b}N_{c}}\mathcal{L}_u $\\
    UpdateStep($\theta$, $\nabla_{\theta}\mathcal{L})$ 
}

 \caption{Semixup}
 \label{alg:semixup}
\end{algorithm}



\begin{table}[H]
  \centering
  \renewcommand{\arraystretch}{1.5}
  \caption{Detailed description of our architecture.}
    \begin{tabular}{lcl}
    \toprule
    \multicolumn{1}{c}{\textbf{Layer name}} & \textbf{Output shape} & \multicolumn{1}{c}{\textbf{Layer}} \\
    \hline \hline
    Input & 128x128 &  \\
    \hline
    Conv1\_1 & 128x128 & ConvBlock 3x3, 32, S = 1 \\
    Conv1\_2 & 128x128 & ConvBlock 3x3, 32, S = 1 \\
    Conv1\_3 & 128x128 & ConvBlock 3x3, 32, S = 1 \\
    Dropout\_1 & 128x128 & Dropout \\
    Conv2\_1 & 64x64 & ConvBlock 3x3, 64, S = 2 \\
    Conv2\_2 & 64x64 & ConvBlock 3x3, 64, S = 1 \\
    Dropout\_2 & 64x64 & Dropout \\
    Conv3\_1 & 32x32 & ConvBlock 3x3, 128, S = 2 \\
    Conv3\_2 & 32x32 & ConvBlock 3x3, 128, S = 1 \\
    Dropout\_3 & 32x32 & Dropout \\
    Conv4\_1 & 16x16 & ConvBlock 3x3, 256, S = 2 \\
    Conv4\_2 & 16x16 & ConvBlock 3x3, 256, S = 1 \\
    Dropout\_4 & 16x16 & Dropout \\
    SAM   & 1x1   & SAM block \\
    \hline
    Merge & 512   & Concat \\
    Dropout\_5 & 512 & Dropout \\
    Output & 5     & Linear, Softmax \\
    \bottomrule
    \end{tabular}%
  \label{tbl:detailed_arch}%
\end{table}%

\begin{figure}[ht]
    \centering
    \IfFileExists{figs/roc_auc/ROC_AUC_50-crop.pdf}{}{\immediate\write18{pdfcrop figs/roc_auc/ROC_AUC_50.pdf}}
    \IfFileExists{figs/roc_auc/ROC_AUC_100-crop.pdf}{}{\immediate\write18{pdfcrop figs/roc_auc/ROC_AUC_100.pdf}}
    \IfFileExists{figs/roc_auc/ROC_AUC_500-crop.pdf}{}{\immediate\write18{pdfcrop figs/roc_auc/ROC_AUC_500.pdf}}
    \IfFileExists{figs/roc_auc/ROC_AUC_1000-crop.pdf}{}{\immediate\write18{pdfcrop figs/roc_auc/ROC_AUC_1000.pdf}}
        \hspace*{\fill}%
        \subfloat[50 labels / KL grade \label{fig:roc50}]{\includegraphics[scale=0.45]{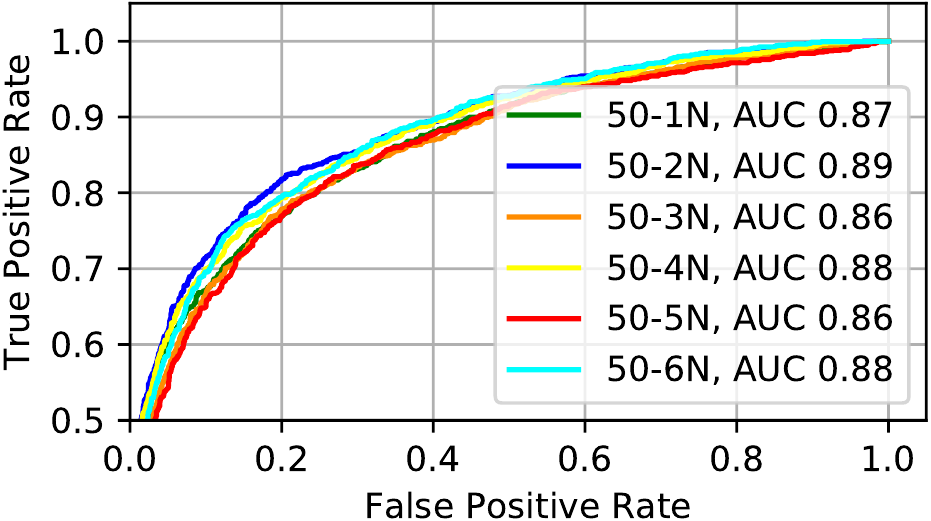}}\hfill%
        \subfloat[100 labels / KL grade \label{fig:roc100}]{\includegraphics[scale=0.45]{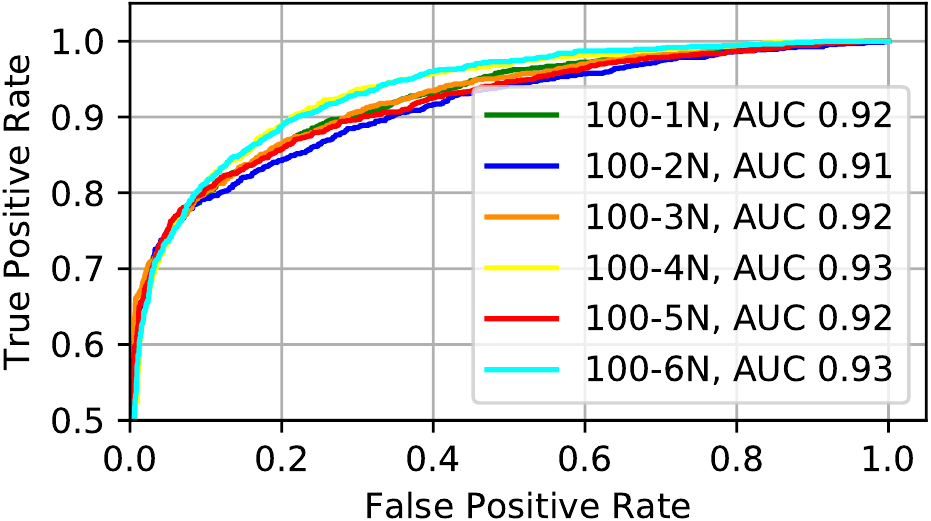}}\hspace*{\fill}%
        \newline \newline
        \hspace*{\fill}%
        \subfloat[500 labels / KL grade
        \label{fig:roc500}]{\includegraphics[scale=0.45]{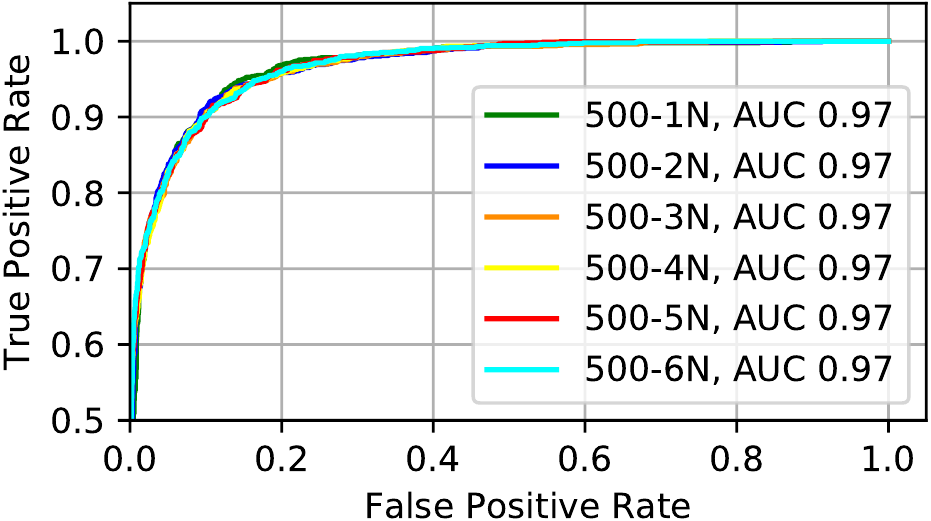}}\hfill%
        \subfloat[1000 labels / KL grade
        \label{fig:roc1000}]{\includegraphics[scale=0.45]{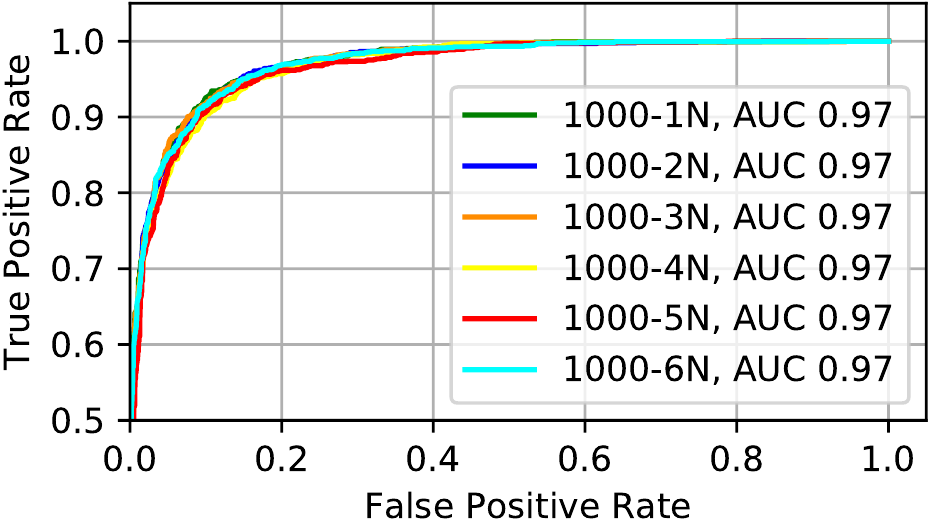}}
        \hspace*{\fill}%
    \caption{ROC curves and AUC of models trained by \textit{Semixup} using $N$ labeled samples per KL grade. Each subplot shows the results of 6 models trained on 6 different amounts of unlabeled data. (\protect\subref*{fig:roc50}) $N=5\times50$. (\protect\subref*{fig:roc100}) $N=5\times 100$. (\protect\subref*{fig:roc500}) $N=5\times500$ (\protect\subref*{fig:roc1000}) $N=5\times1000$.}
    \label{supfig:semixup_roc_auc}
\end{figure}

\begin{table}[H]
    \centering
  \caption{Results (BA means and standard errors,\%) on an independent test set derived from the MOST data for all the \remove{SL and }SSL models \update{with 10 different random seeds}. Here, we varied the amount of labels per KL grade as well as the amount of unlabeled data in each setting (24 settings per each SSL method). The results in bold are the best ones in each  setting. The underline highlights the second best models. The bottom part of the table shows the average performance across all unlabeled data configurations.}\label{suptab:results-all}
  \scalebox{.85}{
    \begin{tabular}{lccccc}
\toprule 
         \multirow{2}{*}{\textbf{Method}} & \textbf{\# unlabeled} & \multicolumn{4}{c}{\textbf{\# labels / KL grade ($N / 5$)}} \\ 
         \cmidrule{3-6}
          & \textbf{data} & \textbf{50} & \textbf{100} & \textbf{500} & \textbf{1000} \\ 
         \midrule \midrule 
 ICT~\cite{verma2019interpolation} & \multirow{4}{*}{$1N$} & 42.9$\pm$0.4 & 48.9$\pm$1.2 & 63.2$\pm$0.6 & 65.9$\pm$0.6\\ 
$\Pi$ model~\cite{laine2016temporal} &  & \underline{44.2$\pm$0.7} & 54.5$\pm$0.7 & 64.3$\pm$0.5 & 65.3$\pm$0.4\\ 
MixMatch~\cite{berthelot2019mixmatch} &  & 41.8$\pm$1.1 & \underline{54.5$\pm$0.6} & \underline{64.7$\pm$0.3} & \underline{66.0$\pm$0.6}\\ 
\textit{Semixup} (Ours) &  & \textbf{46.3$\pm$0.9} & \textbf{56.3$\pm$1.5} & \textbf{66.4$\pm$0.5} & \textbf{67.5$\pm$0.5}\\ 
\midrule 
ICT~\cite{verma2019interpolation} & \multirow{4}{*}{$2N$} & 41.5$\pm$0.8 & 48.7$\pm$0.9 & 63.3$\pm$0.6 & 65.2$\pm$0.4\\ 
$\Pi$ model~\cite{laine2016temporal} &  & \textbf{44.2$\pm$0.5} & 54.0$\pm$0.8 & 64.0$\pm$0.6 & \underline{65.5$\pm$0.5}\\ 
MixMatch~\cite{berthelot2019mixmatch} &  & 40.2$\pm$1.3 & \underline{54.8$\pm$0.6} & \underline{65.1$\pm$0.6} & 65.4$\pm$0.7\\ 
\textit{Semixup} (Ours) &  & \underline{42.9$\pm$3.5} & \textbf{56.2$\pm$1.2} & \textbf{66.1$\pm$0.5} & \textbf{66.1$\pm$0.4}\\ 
\midrule 
ICT~\cite{verma2019interpolation} & \multirow{4}{*}{$3N$} & \underline{45.0$\pm$1.5} & 50.4$\pm$0.8 & 64.0$\pm$0.6 & 65.6$\pm$0.4\\ 
$\Pi$ model~\cite{laine2016temporal} &  & \textbf{45.1$\pm$0.4} & \underline{54.7$\pm$0.5} & \underline{65.2$\pm$0.6} & 65.8$\pm$0.6\\ 
MixMatch~\cite{berthelot2019mixmatch} &  & 42.0$\pm$1.1 & 52.9$\pm$1.3 & 64.6$\pm$0.3 & \underline{66.4$\pm$0.4}\\ 
\textit{Semixup} (Ours) &  & 43.5$\pm$2.9 & \textbf{57.6$\pm$0.7} & \textbf{66.3$\pm$0.6} & \textbf{67.3$\pm$0.4}\\ 
\midrule 
ICT~\cite{verma2019interpolation} & \multirow{4}{*}{$4N$} & 43.9$\pm$0.8 & 50.0$\pm$1.1 & 63.4$\pm$0.4 & 64.8$\pm$0.4\\ 
$\Pi$ model~\cite{laine2016temporal} &  & \underline{45.1$\pm$0.7} & \underline{55.3$\pm$0.6} & 64.1$\pm$0.5 & \underline{65.9$\pm$0.4}\\ 
MixMatch~\cite{berthelot2019mixmatch} &  & 43.3$\pm$0.9 & 53.9$\pm$0.6 & \underline{65.0$\pm$0.5} & 65.8$\pm$0.3\\ 
\textit{Semixup} (Ours) &  & \textbf{48.2$\pm$0.8} & \textbf{57.9$\pm$0.6} & \textbf{65.9$\pm$0.5} & \textbf{66.1$\pm$0.4}\\ 
\midrule 
ICT~\cite{verma2019interpolation} & \multirow{4}{*}{$5N$} & \underline{44.7$\pm$1.4} & 50.9$\pm$0.9 & 63.0$\pm$0.7 & 64.9$\pm$0.5\\ 
$\Pi$ model~\cite{laine2016temporal} &  & \textbf{45.8$\pm$0.5} & \underline{53.1$\pm$0.7} & \underline{64.4$\pm$0.6} & 64.2$\pm$0.4\\ 
MixMatch~\cite{berthelot2019mixmatch} &  & 41.7$\pm$0.8 & 53.1$\pm$0.9 & 63.2$\pm$0.4 & \underline{65.2$\pm$0.2}\\ 
\textit{Semixup} (Ours) &  & 44.4$\pm$2.5 & \textbf{58.4$\pm$0.5} & \textbf{66.0$\pm$0.5} & \textbf{66.1$\pm$0.6}\\ 
\midrule 
ICT~\cite{verma2019interpolation} & \multirow{4}{*}{$6N$} & 44.5$\pm$1.0 & 52.3$\pm$0.7 & 63.3$\pm$0.7 & 64.8$\pm$0.6\\ 
$\Pi$ model~\cite{laine2016temporal} &  & \underline{45.4$\pm$0.4} & 53.4$\pm$0.5 & \underline{64.4$\pm$0.6} & \underline{66.1$\pm$0.5}\\ 
MixMatch~\cite{berthelot2019mixmatch} &  & 41.0$\pm$1.4 & \underline{53.9$\pm$1.0} & 64.3$\pm$0.5 & 65.8$\pm$0.4\\ 
\textit{Semixup} (Ours) &  & \textbf{46.3$\pm$1.5} & \textbf{58.1$\pm$0.8} & \textbf{66.2$\pm$0.6} & \textbf{67.0$\pm$0.5}\\ 
\bottomrule 
\end{tabular}%
}
  \label{tbl:detailed_comparison}%
\end{table}%

\begin{figure}[H]
    \centering
    \IfFileExists{figs/pre_rec/PRE_REC_50-crop.pdf}{}{\immediate\write18{pdfcrop figs/pre_rec/PRE_REC_50.pdf}}
    \IfFileExists{figs/pre_rec/PRE_REC_100-crop.pdf}{}{\immediate\write18{pdfcrop figs/pre_rec/PRE_REC_100.pdf}}
    \IfFileExists{figs/pre_rec/PRE_REC_500-crop.pdf}{}{\immediate\write18{pdfcrop figs/pre_rec/PRE_REC_500.pdf}}
    \IfFileExists{figs/pre_rec/PRE_REC_1000-crop.pdf}{}{\immediate\write18{pdfcrop figs/pre_rec/PRE_REC_1000.pdf}}
        \hspace*{\fill}%
        \subfloat[50 labels / KL grade \label{fig:pr50}]{\includegraphics[scale=0.45]{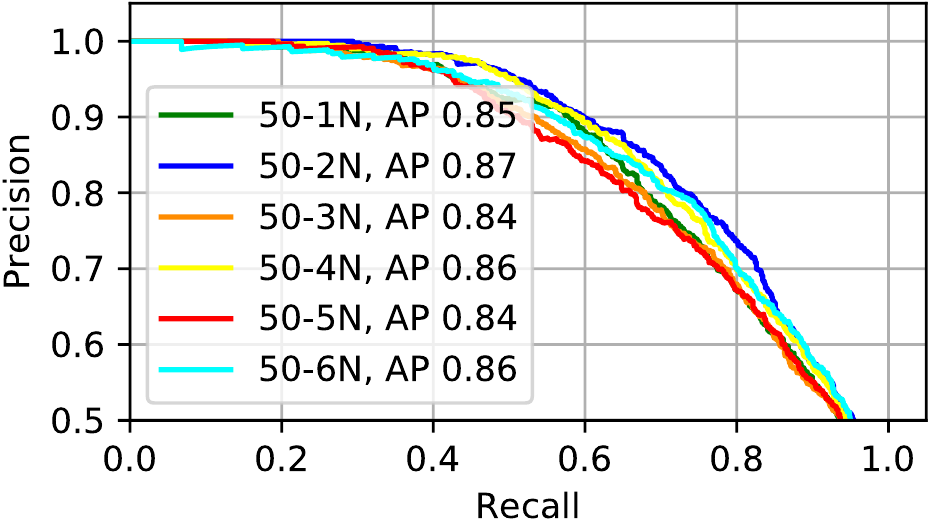}}\hfill%
        \subfloat[100 labels / KL grade \label{fig:pr100}]{\includegraphics[scale=0.45]{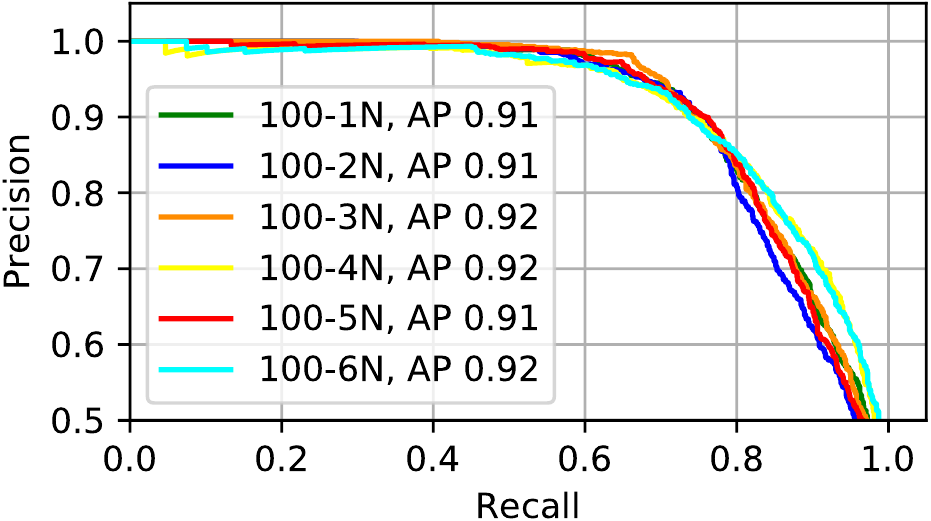}}\hspace*{\fill}%
        \newline \newline
        \hspace*{\fill}%
        \subfloat[500 labels / KL grade
        \label{fig:pr500}]{\includegraphics[scale=0.45]{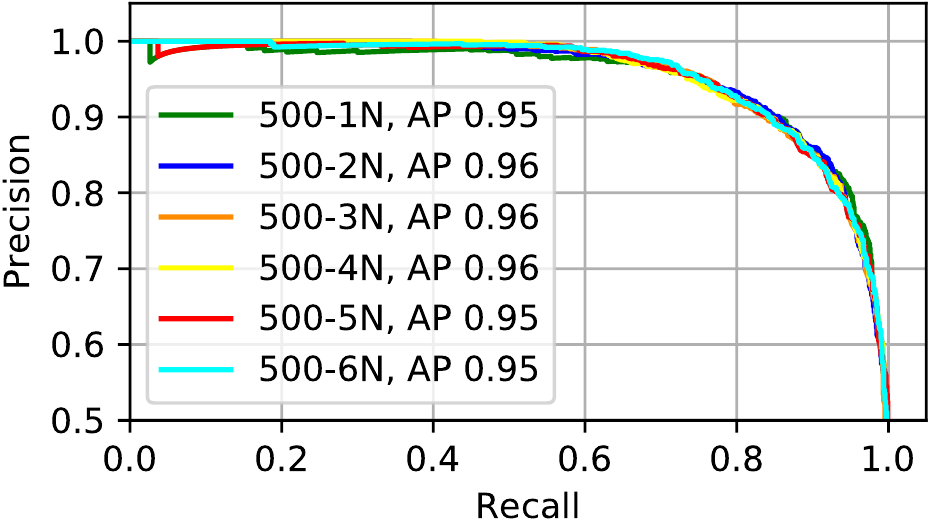}}\hfill%
        \subfloat[1000 labels / KL grade
        \label{fig:pr1000}]{\includegraphics[scale=0.45]{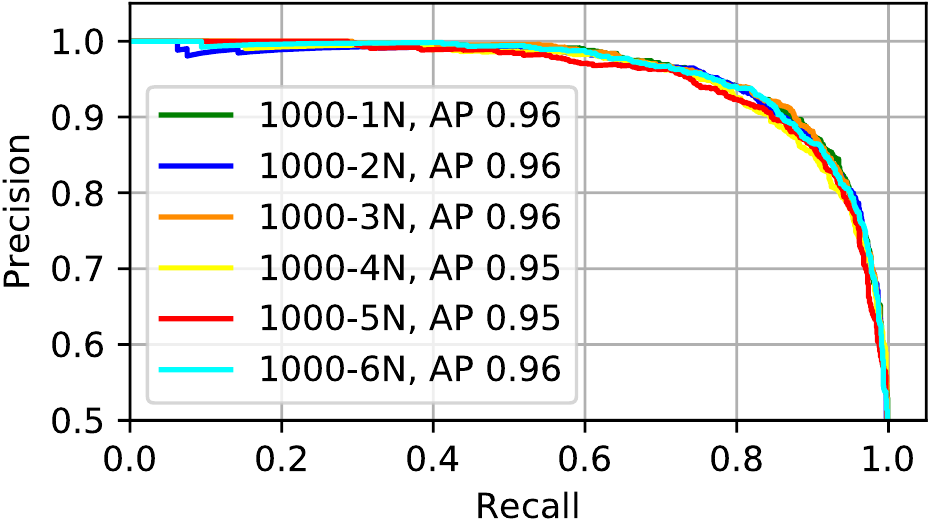}}
        \hspace*{\fill}%
    \caption{PR curves and AP of the models trained by \textit{Semixup} using $N$ labeled samples per KL grade. Each subplot shows the results of 6 models trained on 6 different amounts of unlabeled data. (\protect\subref*{fig:pr50}) $N=5\times50$; (\protect\subref*{fig:pr100}) $N=5\times100$; (\protect\subref*{fig:pr500}) $N=5\times500$; (\protect\subref*{fig:pr1000}) $N=5\times1000$.}
    \label{supfig:semixup_pre_rec}
\end{figure}

\begin{table}[htbp]
  \centering
  \caption{Statistical comparisons using one-sided Wilcoxon signed-rank test. SL$^+$ indicates our fully SL models equipped with either SAM-HV or SAM-VH. SL$^\ddagger$ indicates our model with SAM-HV. * and ** signs correspond to $p < 0.05$ and $p < 0.001$, respectively.}
    \begin{tabular}{crlrrl}
    \toprule
    \multicolumn{2}{c}{\textbf{Main method}} & \multicolumn{2}{c}{\textbf{Compared method}} & \multicolumn{1}{c}{\multirow{2}{*}{\textbf{p-value}}} & \multirow{2}[4]{*}{} \\
\cmidrule{1-2} \cmidrule{3-4}   \textbf{Name} & \multicolumn{1}{c}{\textbf{\# labels}} & \multicolumn{1}{c}{\textbf{Name}} & \multicolumn{1}{c}{\textbf{\# labels}} &       &  \\
    \midrule
    \midrule
    \multirow{20}{*}{\textit{Semixup}} & \multirow{4}{*}{250} & SL$^+$    & \multirow{4}{*}{250} & 5.4e-2  &  \\
          &       & $\Pi$ model~\cite{laine2016temporal} &       & 1.5e-1  &  \\
          &       & ICT~\cite{verma2019interpolation}   &       & 6.3e-1  &  \\
          &       & MixMatch~\cite{berthelot2019mixmatch} &       & 5.4e-2  &  \\
\cmidrule{2-6}          & \multirow{4}{*}{500} & SL$^+$    & \multirow{4}{*}{500} & 5.2e-5  & ** \\
          &       & $\Pi$ model~\cite{laine2016temporal} &       & 1.6e-3  & * \\
          &       & ICT~\cite{verma2019interpolation}   &       & 1.1e-4  & ** \\
          &       & MixMatch~\cite{berthelot2019mixmatch} &       & 5.2e-5  & ** \\
\cmidrule{2-6}          & \multirow{8}{*}{2,500} & SL$^+$    & \multirow{4}{*}{2,500} & 7.5e-4  & ** \\
          &       & $\Pi$ model~\cite{laine2016temporal} &       & 8.4e-2  &  \\
          &       & ICT~\cite{verma2019interpolation}   &       & 2.6e-4  & ** \\
          &       & MixMatch~\cite{berthelot2019mixmatch} &       & 1.2e-1  &  \\
\cmidrule{3-6}          &       & SL$^+$    & \multirow{4}{*}{5,000} & 3.7e-2  & * \\
          &       & $\Pi$ model~\cite{laine2016temporal} &       & 2.5e-1  &  \\
          &       & ICT~\cite{verma2019interpolation}   &       & 6.8e-2  &  \\
          &       & MixMatch~\cite{berthelot2019mixmatch} &       & 1.6e-1  &  \\
\cmidrule{2-6}          & \multirow{4}{*}{5,000} & SL$^+$    & \multirow{4}{*}{5,000} & 2.0e-3  & * \\
          &       & $\Pi$ model~\cite{laine2016temporal} &       & 2.0e-2  & * \\
          &       & ICT~\cite{verma2019interpolation}   &       & 5.8e-4  & ** \\
          &       & MixMatch~\cite{berthelot2019mixmatch} &       & 2.6e-3  & * \\
    \midrule
    \midrule
    \multirow{2}{*}{SL$^\ddagger$} & \multirow{2}{*}{35,000} & \multirow{2}{*}{\textit{Semixup} (Ours)} & 2,500  & 5.4e-2  &  \\
          &  &  & 5,000  & 3.7e-1  &  \\
    \bottomrule
    \end{tabular}%
  \label{subtbl:detailed_wilcoxon_test}%
\end{table}%

\subsection{More Labeled Data}
\label{sc:more_labeled_data}
\begin{figure}[h]
    \centering
    \IfFileExists{figs/semixup_by_nlabels_1N-crop.pdf}{}{\immediate\write18{pdfcrop figs/semixup_by_nlabels_1N.pdf}}
    \IfFileExists{figs/semixup_by_nlabels_2N-crop.pdf}{}{\immediate\write18{pdfcrop figs/semixup_by_nlabels_2N.pdf}}
        \subfloat[$M=N$ \label{fig:semix_n1}]{\includegraphics[scale=.52]{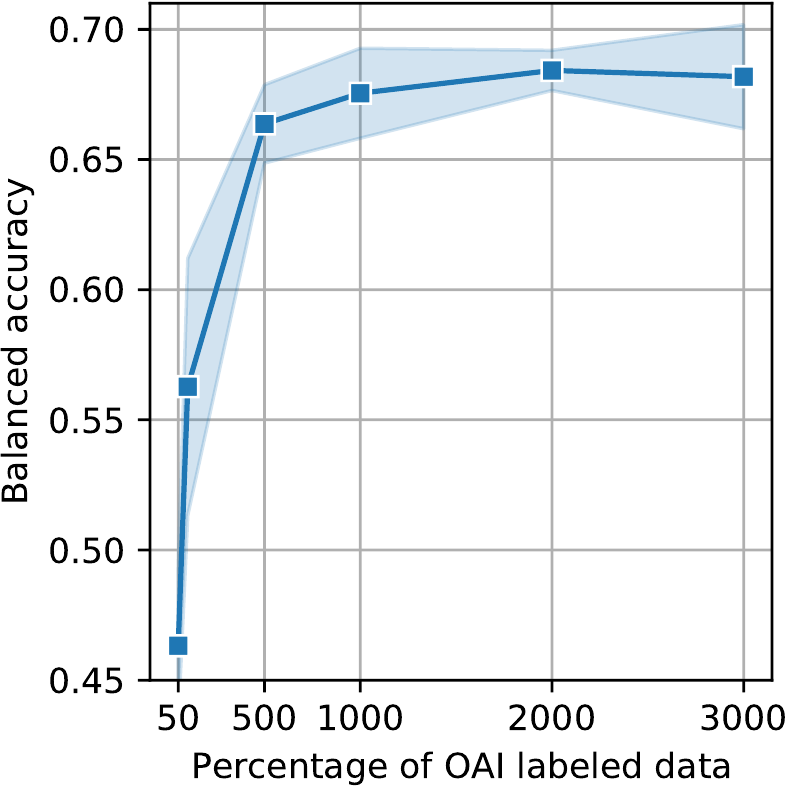}}
        \subfloat[$M=2N$ \label{fig:semix_n2}]{\includegraphics[scale=.52]{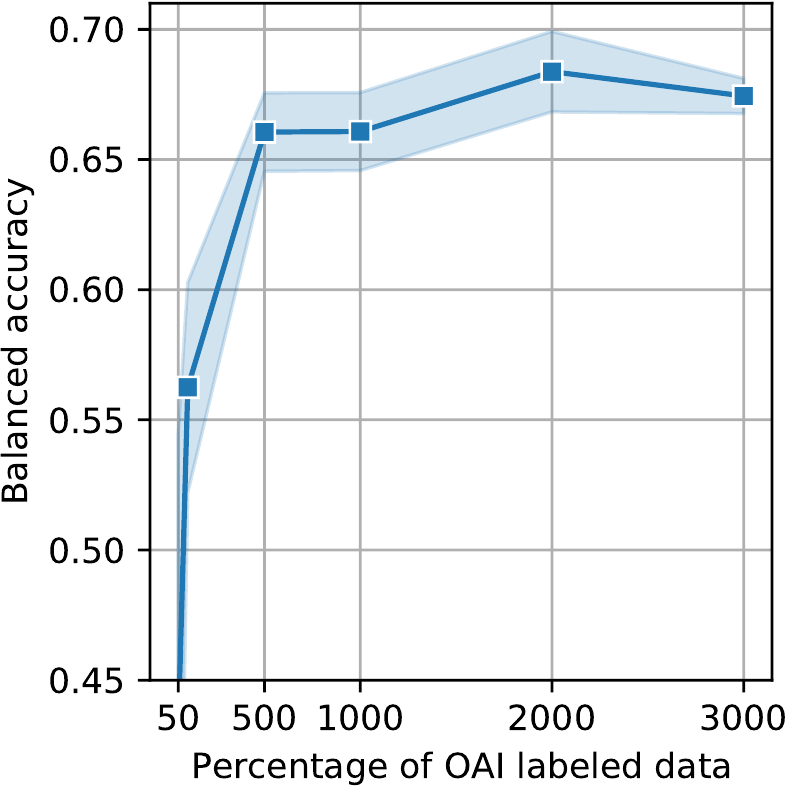}}
    \caption{\small \update{Semixup performance on the MOST with different amounts, $N$, of labeled data. $M$ is the number of unlabeled samples. We run with 10 different random seeds.}}
    \label{fig:semixup_by_nlabels}
\end{figure}

\update{In this section, we extend the number of labeled samples to assess how much Semixup could improve. In detail, we added 2 more data settings that had $2,000$ and $3,000$ labels per KL grade. In the latter setting, the highest feasible number of unlabeled data $M$ is $30,000$ (i.e., $2N$) since it uses all valid OAI data. Hence, we merely conducted 2 comparisons throughout different amounts of labeled data as in \Cref{fig:semixup_by_nlabels}. On the independent test set, we observed that the data setting of $2,000$ labels per KL grade helped to gain a slightly improvement, while the another setting tended to be saturated or even decrease the performance.}

\begin{figure}[h]
    \centering
    \IfFileExists{figs/semixup_data_selection-crop.pdf}{}{\immediate\write18{pdfcrop figs/semixup_data_selection.pdf}}
        \includegraphics[scale=.68]{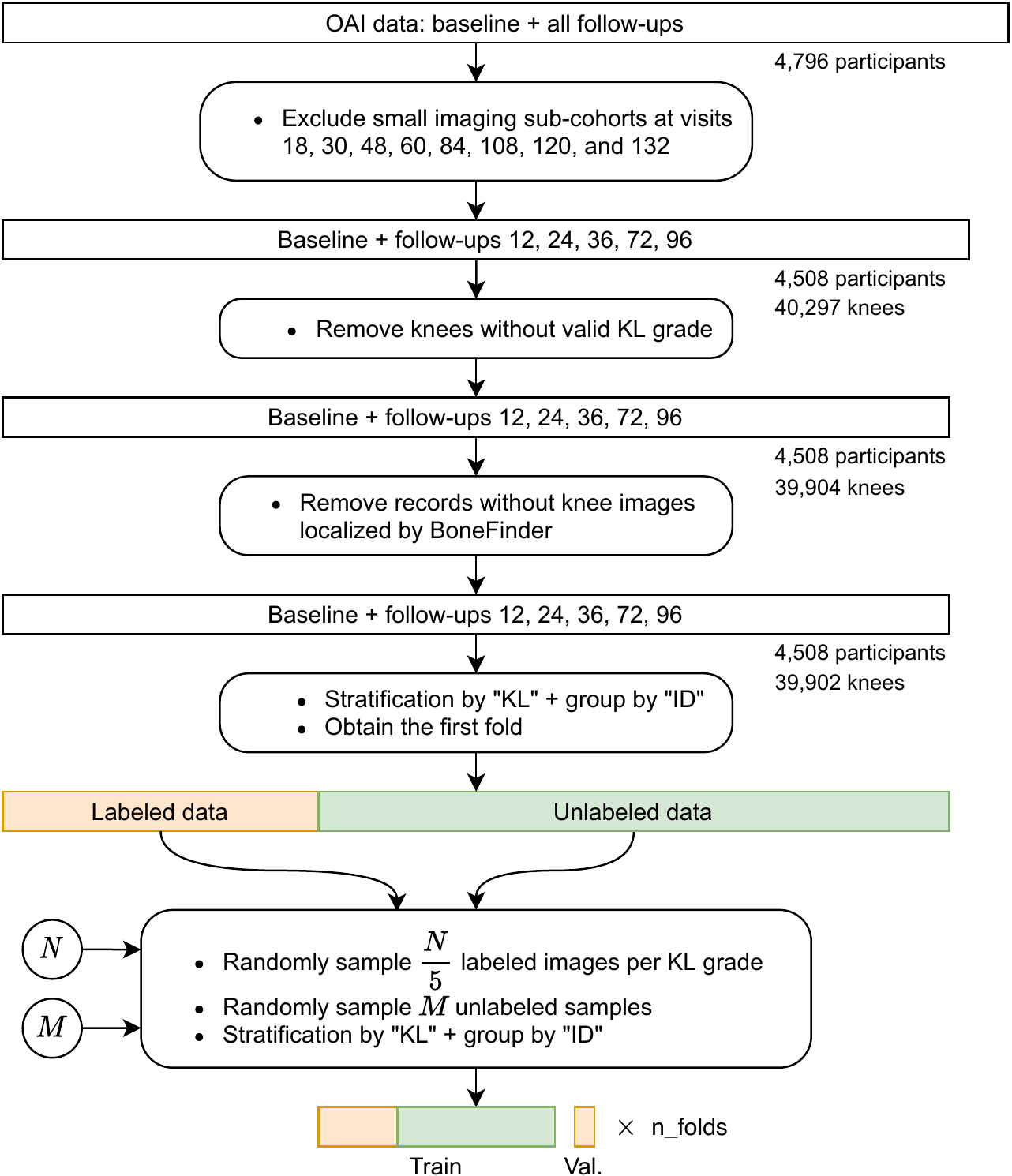}
    \caption{\small \update{Workflow for splitting OAI data.}}
    \label{fig:oai_splitting_workflow}
\end{figure}

\end{document}